\documentclass[]{aa}

\usepackage{amsmath}
\usepackage[dvips]{graphicx}
\usepackage{natbib}
\bibpunct{(}{)}{;}{a}{}{,}
\usepackage[american]{babel}
\usepackage{txfonts}

\usepackage{xspace}

\newcommand{\msol}{\ensuremath{\mathrm{M}_{\odot}}}

\newcommand{\gra}{\raisebox{1ex}{\scriptsize o}}

\def\ga{\mathrel{\mathpalette\fun >}}
\def\fun#1#2{\lower3.6pt\vbox{\baselineskip0pt\lineskip.9pt
  \ialign{$\mathsurround=0pt#1\hfil##\hfil$\crcr#2\crcr\sim\crcr}}}

{\catcode`\|=\active
  \gdef\Braket#1{\left<\mathcode`\|"8000\let|\bravert {#1}\right>}}
\def\bravert{\egroup\,\vrule\,\bgroup}

\def\gtrsim{\ensuremath{\; \buildrel > \over \sim \;}}
\def\gsim{\lower.5ex\hbox{\gtsima}}

\newcommand{\sax}{\textsl{BeppoSAX}\xspace}
\newcommand{\ssax}{\textsl{SAX}\xspace}

\newcommand{\mecs}{\textsl{MECS}\xspace}
\newcommand{\pds}{\textsl{PDS}\xspace}
\newcommand{\hp}{\textsl{HPGSPC}\xspace}
\newcommand{\rxte}{\textsl{RXTE}\xspace}
\newcommand{\pca}{\textsl{PCA}\xspace}
\newcommand{\hexte}{\textsl{HEXTE}\xspace}
\newcommand{\hextea}{\textsl{HEXTE-A}\xspace}
\newcommand{\hexteb}{\textsl{HEXTE-B}\xspace}

\newcommand{\prog}[1]{\texttt{#1}\xspace}
\newcommand{\xspec}{\prog{XSPEC}\xspace}
\newfont{\mc}{cmcsc10 scaled\magstep2}
\newfont{\cmc}{cmcsc10 scaled\magstep1}

\newcommand{\eflux}{{\rm erg\,cm^{-2}\,s^{-1}}}
\newcommand{\fu}{4U~0115$+$63\xspace}
\newcommand{\her}{Her~X-1\xspace}
\newcommand{\cen}{Cen~X-3\xspace}

\newcommand{\bgi}{\begin{itemize}}
\newcommand{\edi}{\end{itemize}}

\newcommand{\be}{\begin{equation}}
\newcommand{\ee}{\end{equation}}

\newcommand{\bea}{\begin{eqnarray}}                  %
\newcommand{\eea}{\end{eqnarray}}                    %

\newcommand{\beaa}{\begin{eqnarray*}}                %
\newcommand{\eeaa}{\end{eqnarray*}}                  %

\newcommand{\bgd}{\begin{description}}
\newcommand{\edd}{\end{description}}

\newcommand{\bgf}{\begin{figure}}
\newcommand{\edf}{\end{figure}}

\newcommand{\bgc}{\begin{center}}
\newcommand{\edc}{\end{center}}

\newcommand{\bgt}{\begin{tabular}}
\newcommand{\edt}{\end{tabular}}

\newcommand{\bge}{\begin{enumerate}}
\newcommand{\ede}{\end{enumerate}}

\DeclareRobustCommand{\ion}[2]{%
 \relax\ifmmode
 \ifx\testbx\f@series
 {\mathbf{#1\,\mathsc{#2}}}\else
 {\mathrm{#1\,\mathsc{#2}}}\fi
 \else\textup{#1\,{\mdseries\textsc{#2}}}%
 \fi}

\def\jnl@style{\it}
\def\aaref@jnl#1{{\jnl@style#1}}

\def\aaref@jnl#1{{\jnl@style#1}}

\def\aj{\aaref@jnl{AJ}}                   
\def\araa{\aaref@jnl{ARA\&A}}             
\def\apj{\aaref@jnl{ApJ}}                 
\def\apjl{\aaref@jnl{ApJ}}                
\def\apjs{\aaref@jnl{ApJS}}               
\def\ao{\aaref@jnl{Appl.~Opt.}}           
\def\apss{\aaref@jnl{Ap\&SS}}             
\def\aap{\aaref@jnl{A\&A}}                
\def\aapr{\aaref@jnl{A\&A~Rev.}}          
\def\aaps{\aaref@jnl{A\&AS}}              
\def\azh{\aaref@jnl{AZh}}                 
\def\baas{\aaref@jnl{BAAS}}               
\def\jrasc{\aaref@jnl{JRASC}}             
\def\memras{\aaref@jnl{MmRAS}}            
\def\mnras{\aaref@jnl{MNRAS}}             
\def\pra{\aaref@jnl{Phys.~Rev.~A}}        
\def\prb{\aaref@jnl{Phys.~Rev.~B}}        
\def\prc{\aaref@jnl{Phys.~Rev.~C}}        
\def\prd{\aaref@jnl{Phys.~Rev.~D}}        
\def\pre{\aaref@jnl{Phys.~Rev.~E}}        
\def\prl{\aaref@jnl{Phys.~Rev.~Lett.}}    
\def\pasp{\aaref@jnl{PASP}}               
\def\pasj{\aaref@jnl{PASJ}}               
\def\qjras{\aaref@jnl{QJRAS}}             
\def\skytel{\aaref@jnl{S\&T}}             
\def\solphys{\aaref@jnl{Sol.~Phys.}}      
\def\sovast{\aaref@jnl{Soviet~Ast.}}      
\def\ssr{\aaref@jnl{Space~Sci.~Rev.}}     
\def\zap{\aaref@jnl{ZAp}}                 
\def\nat{\aaref@jnl{Nature}}              
\def\iaucirc{\aaref@jnl{IAU~Circ.}}       
\def\aplett{\aaref@jnl{Astrophys.~Lett.}} 
\def\apspr{\aaref@jnl{Astrophys.~Space~Phys.~Res.}}
\def\bain{\aaref@jnl{Bull.~Astron.~Inst.~Netherlands}} 
\def\fcp{\aaref@jnl{Fund.~Cosmic~Phys.}}  
\def\gca{\aaref@jnl{Geochim.~Cosmochim.~Acta}}   
\def\grl{\aaref@jnl{Geophys.~Res.~Lett.}} 
\def\jcp{\aaref@jnl{J.~Chem.~Phys.}}      
\def\jgr{\aaref@jnl{J.~Geophys.~Res.}}    
\def\jqsrt{\aaref@jnl{J.~Quant.~Spec.~Radiat.~Transf.}}
\def\memsai{\aaref@jnl{Mem.~Soc.~Astron.~Italiana}}
\def\nphysa{\aaref@jnl{Nucl.~Phys.~A}}   
\def\physrep{\aaref@jnl{Phys.~Rep.}}   
\def\physscr{\aaref@jnl{Phys.~Scr}}   
\def\planss{\aaref@jnl{Planet.~Space~Sci.}}   
\def\procspie{\aaref@jnl{Proc.~SPIE}}   

\begin{document}

\title{\object{4U\,0115$+$63}: phase lags and cyclotron resonant scattering}

\author{C. Ferrigno
        \inst{1}
        \and
        M. Falanga\inst{2}
        \and 
        E. Bozzo\inst{1}
	\and
      	P. A. Becker\inst{3}
        \and
        D. Klochkov\inst{4}
        \and
        A. Santangelo\inst{4}
      }

\authorrunning{C. Ferrigno et al.}
\titlerunning{Phase lags in \fu}
   \offprints{C. Ferrigno}

\institute{ISDC data center for astrophysics,  Universit\'e de Gen\`eve, chemin d'\'Ecogia, 16 1290 Versoix Switzerland\\
	\email{Carlo.Ferrigno@unige.ch}
	 \and
	 International Space Science Institute (ISSI) Hallerstrasse 6, CH-3012 Bern, Switzerland
	  \and
           Department of Computational and Data Sciences \\George Mason University\\
	   4400 University Drive, MS 6A3 Fairfax, VA 22030
	\and
           IAAT, Abt.\ Astronomie, Universit\"at T\"ubingen,
           Sand 1, 72076 T\"ubingen, Germany
          }

\date{Received ---; accepted ---}

\abstract
	{High-mass X--ray binaries are among the brightest 
         objects of our Galaxy in the high-energy domain (0.1-100\,keV). Despite our relatively good  
       	knowledge of their basic emission mechanisms, the complex problem of understanding their time- and
	energy- dependent X--ray emission
	has not been completely solved.} 
	{In this paper, we study the energy-dependent pulse profiles of the high-mass
         X-ray binary pulsar \fu\ to investigate how they are affected by
         cyclotron resonant scattering. } 
         {We analyzed archival \sax and \rxte observations performed during 
         the giant outburst of the source that occurred in 1999. 
         We exploited a cross correlation technique to compare the pulse profiles 
         in different energy ranges and developed a relativistic ray-tracing model to interpret our findings. We also studied the phase
         dependency of the cyclotron absorption features by performing phase-resolved spectroscopy.} 
        {The pulse profiles of \fu\ displayed clear ``phase-lags'' at energies close to those 
        of the cyclotron absorption features that characterize  
        the X-ray emission of the source. We reproduce this phenomenon qualitatively by assuming an energy-dependent
        beaming of the emission from the column surface and verify that our model is also compatible with the results of 
        phase-resolved spectral analysis.} 
        {We showed that cyclotron resonant scattering affects the pulse profile formation mechanisms in a 
	complex way, which necessitates both improvements in the modeling and the study of  
	other sources to be better understood.
	} 


\maketitle

\section{Introduction}
\label{sec:intro} 

X-ray binary pulsars (XRBPs) are among the brightest
objects in X-rays, and they were discovered 
in the early 70s thanks to the pioneering observations 
carried out by \citet{giac71} and \citet{tananbaum1972}. 
Many XRBPs comprise a young neutron star (NS), endowed with a strong 
magnetic field ($B\sim10^{12}$\,G) and a supergiant or main sequence ``donor'' star. 
Depending on its nature, the donor star can lose a conspicuous amount of 
matter through a strong stellar wind and/or via 
the Roche lobe overflow \citep[see e.g.,][]{frank2002}.  
A large part of this material is focused toward the NS as a consequence of the   
strong gravitational field of this object, and then threaded by its intense magnetic field 
at several thousand kilometers from the stellar surface. Once funneled 
down to the magnetic poles of the NS, accretion columns may form. 
Here, the gravitational potential energy 
of the accreting matter is first converted into kinetic energy 
and then dissipated in the form of X-rays.
The luminosity generated by this process can reach values of  
$\sim$10$^{38}$~erg/s in the energy band 0.1-100~keV \citep[see e.g.,][]
{Pringle72,Davids73}. To a first approximation, 
the magnetic field lines of the NS rotate rigidly with the surface of the star, 
and thus the X-ray emission emerging from the accretion column 
is received by the observer 
modulated at the spin period of the compact object, 
if the magnetic and rotational axes are misaligned.

When averaged over many rotational cycles, the
X-ray emission of the XRBPs folded at the spin period 
of the NS, i.e. the so-called ``pulse profile'', 
shows a remarkable stability.
Significant variations 
in the shape of profiles are observed when the XRBPs undergo transitions between different X-ray luminosity states, such
as reported in the cases of EXO\,2030$+$275  and V\,0332$+$65, \citep[][ and references therein]{dima2008,tsygankov2010}.  
They have, so far, been relatively well understood in terms of changes in the geometry of the 
intrinsic beam pattern\footnote{In this paper, the ``\emph{intrinsic beam pattern}'' is the
intensity of radiation as a function of its orientation 
with respect to the infinitesimal surface element on the column surface, measured in the frame comoving with the column. 
The ``\emph{asymptotic beam pattern}'' or ``\emph{beam pattern}'' is
an estimate of the flux integrated along the column surface as a function of the angle formed by the column
axis with respect to
the line of sight to the observer. It is measured at the location of the observer, and can be obtained from the intrinsic 
beam pattern after accounting for the general relativistic effects.
Only the pulse profile can be measured directly from the observations.}. 
At higher luminosities than the critical value for the characteristics of \fu
\citep[$L_{\rm X}\gtrsim4\times10^{37}\,\mathrm{erg\,s^{-1}}$,][]{basko1976}, a radiative shock forms 
in a relatively extended (a few km) accretion column, and the radiation is emitted mainly from its 
lateral walls in the form of a ``fan beam''.  At lower luminosities, the radiative shock is suppressed, and matter
reaches the base of the column with the free-fall velocity. 
Thus, the height of the column can be significantly reduced, and
for very weak sources, it reduces to a hot-spot on the NS surface. In this case,  
the X-ray radiation escapes mostly along the axis of the column and the system shows a typical 
``pencil beam''. When the X-ray luminosity of a source is 
close to the Eddington limit a combination of pencil and fan beams might apply 
\citep[as proposed for Her~X-1,][]{leahy2004a,leahy2004b}. In some cases, it has also been shown that 
the emission can originate in hollow cones \citep[][ and references therein]{kraus2001,leahy2003}, be reprocessed 
onto the NS surface to produce a soft scattering halo around the base of the column, and/or be shadowed by the
upper part of the accretion stream \citep{kraus1989,kraus2003}. 

The majority of XRBPs show significant changes in the spectral energy distribution of the X-ray emission  
at different pulse phases. This is due to a variety of factors that change during 
the rotation of the star. In particular, the cross section of the 
scattering between photons and electrons trapped in a strong magnetic field
strongly depends on the angle between the photon trajectory and the magnetic field lines, but
the occultation of part of the column during the rotation of the 
NS plays also a significant role.
 
Among the different radiation processes, the cyclotron resonant scattering is probably 
the most crucial for the XRBPs, as it gives rise to the cyclotron resonant scattering 
features \citep[CRSFs; see, e.g.,][ for recent models]{isenberg1998,araya2000,gabi2007}. 
These features appear in the spectra of the XRBP sources in the form of  
relatively broad (1-10~keV) absorption lines. The CRSFs, if detected, provide  
a unique tool to directly estimate the NS magnetic field strength, because  
the centroid energy of the fundamental appears at 
$E_{\rm cyc} = 11.6\,B_{12}\times (1+z)^{-1}$\,keV, and  
higher harmonics can be observed at roughly integer multiples of $E_{\rm cyc}$
\citep[Here $B_{12}$ is the magnetic field strength in units of $10^{12}$\,G and $z$ is the
gravitational redshift in the line-forming region,][]{wasserman1983}.  
Furthermore, since the intrinsic beam pattern emerges from the accretion column relatively close to the 
surface of the NS, the general relativity effects related to the ``compactness'' of this object 
(i.e., the ratio between its mass and radius) affects the way this radiation 
is perceived by a distant observer at different rotational phases. 

Despite the intrinsic complexity of the problem, it was soon realized that the joint study of 
the pulsed profiles and spectral properties of the X-ray radiation from the XRBPs at different 
rotational phases would have provided an unprecedented opportunity to investigate 
the physics of these objects and the interaction between matter and radiation in the presence 
of strong gravitational and magnetic fields.   

Early attempts to link the spectral characteristics of the radiation emerging from an accretion 
column to the pulse profiles of the XRBPs were presented by \citet{meszaros1985a,meszaros1985b}.
Taking some of the basic interactions between matter and radiation in the column into account, 
these authors were able to produce the first energy-dependent beam patterns that could be used to 
predict the observed properties of pulse profiles and spectra of the XRBPs. A few years later, \citet{riffert1988a},  
\citet{riffert1988b}, and \citet{leahy1995} improved these calculations by including 
most of the relevant relativistic effects: light bending, gravitational red shift, and special relativistic beaming near the compact object. These are essential for comparing the model with 
the real observed pulse profiles \citep{riffert1993}.  

The major problem with these models is the symmetry of the simulated 
pulse profiles, whereas the real ones 
are asymmetric and characterized by complex structures. 
The symmetry is mainly due to the simplifications introduced in the theoretical treatment 
of the accretion process and the magnetic field geometry. A number of attempts to use different 
geometries have been discussed by \citet{leahy1990,leahy1991}, and \citet{riffert1993}, 
but they were only partially successful. The most relevant improvement 
was obtained by introducing a misalignment between the two magnetic poles on 
the NS surface. 

As all the physical processes and the geometrical complications described above can hardly be 
taken into account within a single theoretical model, \citet{kraus1995} propose
an alternative and promising way to investigate the pulse profile formation mechanism by 
means of a pulse decomposition method.
From a set of observed pulse profiles in several energy bands and luminosity states, it is possible
to determine the location of the accretion columns on the NS and thus decouple the geometrical effects in a 
nearly model-independent way\footnote{The basic assumption is the equality of the beam patterns 
of the two accretion columns. Moreover, the decomposition is unique only if the angle between the rotation axis 
and the line of sight is known, otherwise this angle must be assumed.}. 
The method was successfully applied to a number of XRBPs for which 
observations with sufficiently high statistics were available: \cen\ \citep{kraus1996}, 
\her\ \citep{blum2000}, \object{EXO\,2030$+$275} \citep{sasaki2010}, \object{A\,0535$+$26} \citep{caballero2010}, 
\object{V\,0332$+$53}, and \fu\ \citep{sasaki2011}. The reconstructed beam patterns
show that, for EXO\,2030$+$275 and \fu, the beam patterns could be 
qualitatively modeled as the combination of the emission from a filled column 
(fan and/or pencil beam depending on luminosity), 
plus a soft scattering halo, whereas for A\,0535$+$26 and V\,0332$+$53
the presence of a hollow column seems to be preferred. 

The reconstruction method presented by \citet{kraus1995} has so far proved
to be able to unveil some of the main geometrical properties of the accretion process in 
the brightest XRBPs. However, the beam patterns have only
been reconstructed for rather large energy intervals (several keV) and thus they cannot 
be used to investigate how effects occurring in narrow energy ranges, e.g., the CRSFs, 
can affect the beam patterns.

In this work, we begin a study of the energy dependency of the pulse profile in XRBPs, at energies 
close to that of the CRSFs. As a case study, we consider the source \fu,\ so far the only XRBPs that 
displayed CRSFs up to the sixth harmonics \citep{heindl2004,ferrigno2009}. 
A summary of the known properties of this source is provided in Sect.~\ref{sec:fu}. 
In Sect.~\ref{sec:observations}, we describe the data analysis technique and the results 
of the \rxte\ and \sax\ observations of \fu\ performed during the giant 
outburst of the source in 1999. 
We find clear evidence for phase shifts in the pulse profile of \fu\ 
at energies close to those of the different harmonics of the CRSF. 
In Sect~\ref{sec:model}, we propose a simplified model to interpret these shifts 
in terms of energy dependent variations of the intrinsic beam pattern 
(see also Appendix \ref{sec:appendix}). We discuss our most relevant 
results and draw our conclusions in Sect.~\ref{sec:discussion}.

\section{\fu}
\label{sec:fu}

\fu\ was discovered in 1969 during one of its giant type II 
X-ray outbursts \citep{whitlock1989}. On that occasion the 
X-ray luminosity of the source rose to $\gtrsim$10$^{37}$\,erg/s, 
a value that is about two orders of magnitude higher than its  
typical quiescent level. X-ray pulsations at a period of 
$P_S=3.6$\,s were reported  
first by \citet{cominski1978}. \citet{rappaport1978} measured later 
the orbital parameters of the system by using 
\textsl{SAS} data, and estimated an orbital period 
of $P_\mathrm{orb}=24.3\,\mathrm{d}$, an eccentricity of $e=0.34$, 
and a value of the semi-major axis of the orbit of $a_X \sin i =
140.1\,\mathrm{lt-s}$. Here $i$ is the unknown inclination angle 
between the normal to the plane of the orbit and the line of 
sight to the observer. 

Optical and IR observations permitted  the object 
V~635~Cas to be identified as the companion star of \fu\ \citep{johns1978}, located  
at a distance of 7-8\,kpc \citep{negueruela2001a}.
An in-depth study of the relation between optical and X-ray emissions  
from the system showed that the type II outbursts from \fu\ 
were occurring almost regularly every three years or so \citep{whitlock1989}, 
and were most likely related to instabilities in the disk due to radiative warping 
\citep{negueruela2001a,negueruela2001b}.  

In the hard X-ray domain, the continuum energy distribution from \fu\ could be reasonably well 
modeled by using, e.g., the phenomenological model NPEX  
\citep{nakajima2006}.
Cyclotron absorption lines were observed in the spectra of this source
up the sixth harmonics \citep{wheaton1979,white1983,heindl1999,santangelo1999,heindl2004,ferrigno2009}. 
The centroid energy of the fundamental line is usually 
at $\sim$11--16\,keV, the lowest among those measured from the other Galactic accreting NS 
displaying evidence of cyclotron absorption lines.
The analysis of the data obtained with the \textsl{GINGA} satellite revealed that 
both the continuum emission from the source and the 
centroid energies of the different cyclotron lines were changing significantly 
with the NS pulse phase \citep{mihara2004}. 
By using \textsl{RXTE} data, \citet{nakajima2006} show that 
the energy of the lines also depends on the  
source overall luminosity, and suggest that  
these variations are related to the transition from 
a radiation-dominated flow  at high luminosities (with a shock above the 
stellar surface) to a matter-dominated flow at low luminosities.  
According to this interpretation, the authors provide the first 
constraints on the height of the accretion column where the cyclotron lines 
are generated ($\sim$0.2-0.3 times the radius of the NS). 
\citet{russi2007}. They confirm that the height of the accretion column 
decreases at lower luminosities and provide constraints on the size of
the line forming region.

The first detailed account of the broad-band, phase-resolved spectral properties of the X-ray emission 
from \fu\ based on a physical model has been reported by \citet{ferrigno2009}. These authors apply it to  
one \sax data set, collected during the giant outburst of the source in 1999, the 
self-consistent model developed by \citet{becker2007} to interpret the high-energy 
emission from XRBPs. They find that the continuum emission from the source 
at energies lower than the fundamental cyclotron line ($\sim$10~keV) is roughly 
constant through the pulse phase, and this component could be 
described by a model of thermal Comptonization of a blackbody with a temperature of 0.5\,keV 
and a radius comparable to the size of the NS ($\sim$10\,km).
Conversely, the spectral properties of the emission above $\sim$10\,keV can be understood in terms of 
Compton reprocessing of the cyclotron emission. The latter is the main cooling channel 
of the accretion column in \fu and has been observed to be more prominent at the peak of the pulse.
These results suggested that the observed radiation from \fu\ during periods of intense X-ray activity 
($L_{\rm X}$$\gtrsim$10$^{37}$\,erg/s) is most likely escaping from the lateral walls of the accretion column 
 and also from a scattering halo close to the surface of the NS.
Interestingly, applying the beam pattern reconstruction method \citep[][see also Sect.~\ref{sec:intro}]
{kraus1995} on the same data also supports this conclusion \citep{sasaki2011} and provides the 
first constraints on the system geometrical properties (see Sect.~\ref{sec:model}). 

\section{Observations and data analysis.}
\label{sec:observations}

For this work, we used the observations of \fu\ performed with \sax\ 
\citep[also indicated hereafter as \ssax,][]{sax} and \rxte\ \citep{rxte} during the giant 
outburst that occurred in 1999. A log of all the observations is provided in Table~\ref{tab:observations}.
Results obtained from these data have already been presented by \citet{santangelo1999}, \citet{heindl1999}, \citet{heindl2004}, \citet{nakajima2006}, and \citet{ferrigno2009}.
\begin{table*}
\caption{Log of the \sax\ and \rxte\ observations of \fu\ used for this work. 
The uncertainty on spin period measurements is at $1\sigma$ c.l. on the last digit. 
The source flux is measured from the spectral analysis and the uncertainty is only 
statistical at 90\% c.l. Assuming a source distance of 7\,kpc, a 
flux of $2\times10^{-8}\,\mathrm{erg\,s^{-1}\,cm^{-2}}$ would 
correspond to $L_{2-100\,\mathrm{keV}}=1.2\times10^{38}\,\mathrm{erg\,s^{-1}}$.} 
\begin{center}
 \begin{tabular}{ l c c c c c c c c}
\hline
\hline
Obs & Instrument   & Start Time [UT] & Stop Time [UT] & \multicolumn{3}{c}{Exposures [ks]} & P [s] & Flux (2-100\,keV)\\
              &   \ssax                &        &           & \mecs&\hp&\pds& &[$10^{-8}\eflux$]\\
              &   \rxte                &        &           &  \pca&\hextea&\hexteb & \\
\hline
(a) & \ssax  &  1999-03-06 14:48 & 1999-03-07 14:23 & 54.9 & 48.8 & 44.3 & 3.61456[2] & $1.61\pm0.01$\\ 
(b) & \rxte  &  1999-03-07 04:09 & 1999-03-07 09:26 &  17.1 & 7.3 & 7.2 &3.614593[6] & $1.70\pm0.02$\\
(c) & \rxte  &  1999-03-12 00:02 & 1999-03-12 07:30 &  13.8& 7.7 & 7.7 & 3.61428[2] & $2.05\pm0.01$\\
(d) & \ssax &  1999-03-19 17:05 & 1999-03-20 08:42 &  31.2 & 32.0 & 30.0 &  3.613816[3]$^*$ & $1.62\pm0.01$\\
(e) & \rxte  &  1999-03-21 06:27 & 1999-03-21 08:33 &  8.8 & 1.5 & 1.4 & 3.61441[8] & $2.05\pm0.04$\\
(f) & \ssax  &  1999-03-26 17:31 & 1999-03-27 17:34 & 53.7 & 42.5 & 48.3 & 3.614246[2] & $1.07\pm0.02$\\
(g) & \rxte  &  1999-03-28 01:04 & 1999-03-28 08:12 &  8.8 & 1.7 & 1.7 & 3.61443[1] & $1.13\pm0.08$\\
(h) & \rxte  &  1999-03-29 04:15 & 1999-03-29 11:17 &  5.9 & 3.5 & 3.5 & 3.61444[1] & $1.04\pm0.06$\\
 \hline
\multicolumn{9}{l}{$^*$ The period is referred to MJD 51\,256.7037037 (TBD), when $\dot P = -(1.56\pm0.06)\times 10^{-9}$\,s/s.}
\end{tabular}
\label{tab:observations}
\end{center}
\end{table*}

\sax\ data were analyzed with LHEASOFT v5.2, for which our research group maintained a working version 
of the \textsl{Saxdas} software.
We considered all the data obtained with the high-energy narrow field instruments   
\mecs \citep[1.5--10.5\,keV,][]{mecs}, \hp \citep[7--44\,keV, ][]{hp}, and \pds \citep[15--100\,keV,
][]{pds}. All data were reprocessed through the standard pipeline
(\texttt{saxdas} v.2.1). Background subtraction was performed using the
method of the earth occultation for the \hp,\ the off-source pointing for the \pds,\ and the
standard calibration files for the \mecs.\ In the first observation, designated (a), 
we lowered the earth occultation threshold of the \hp\ to $-1\gra$ instead of the standard $-2\gra$, 
in order to properly estimate the background. The \mecs\ source extraction radius was set 
to $8^\prime$ to optimize the S/N. 

\rxte\ data were analyzed with heasoft v6.10. We considered data from the \pca\ \citep[2-60\,keV, ][]{pca} 
and from the \hexte\ \citep[15--250\,keV, ][]{hexte}. Good Time Intervals (GTI) were created by imposing that 
at least 30 minutes passed after the satellite exited the SAA, and the elevation on the Earth limb was at least 10\gra.  
For the \pca,\ we additionally requested an electron ratio $\leq$0.15. We modeled the background using the estimator 
v3.8 for bright sources, and extracted the events in the mode \texttt{E\_125us\_64M} (time resolution 122\,$\mu$s 
and 64 energy channels). The instrument specific response matrices were generated considering the corrected 
energy bounds and the gain corrections performed by the \textsl{EDS}.  We estimated the background for the \hexte\ 
instrument from the available off-set pointing and generated the corresponding response matrix with standard 
procedures.

\subsection{Timing analysis}
\label{sec:timing}

The timing analysis was performed exploiting the best known position of \fu\ 
(Ra, Dec=19.630,63.740~deg, J2000) in the most suitable energy ranges 
for the different instruments on-board \sax\ and  \rxte\ 
 (1.65--8.5\,keV, \mecs;\ 8.5--25\,keV, \hp;\ 25--100\,keV, \pds;\ 3--45 keV, \pca;\ 45--80\,keV, \hexte\ ).   
All the photon arrival times were first converted to the Solar system barycenter with the tools 
\texttt{baryconv} (\sax) or \texttt{faxbary} (\rxte), and then calculated in the 
center of mass of the binary using the available orbital ephemeris of \fu\ 
\citep{tamura1992}. 
In each observation, the source spin period was estimated with a phase-shift method 
\citep[see e.g. ][]{ferrigno2007}, and the results are reported in Table~\ref{tab:observations}.  
The relatively high uncertainty affecting the period measurements is mainly due to the timing noise 
related to the variation of the pulse profile from one pulse to the other. This produces a 
scatter of $\sim$2\% on the phase of the first and second Fourier harmonics used in this 
analysis. 
\begin{figure}
  \begin{center}
  \resizebox{\hsize}{!}{
      \includegraphics[angle=0]{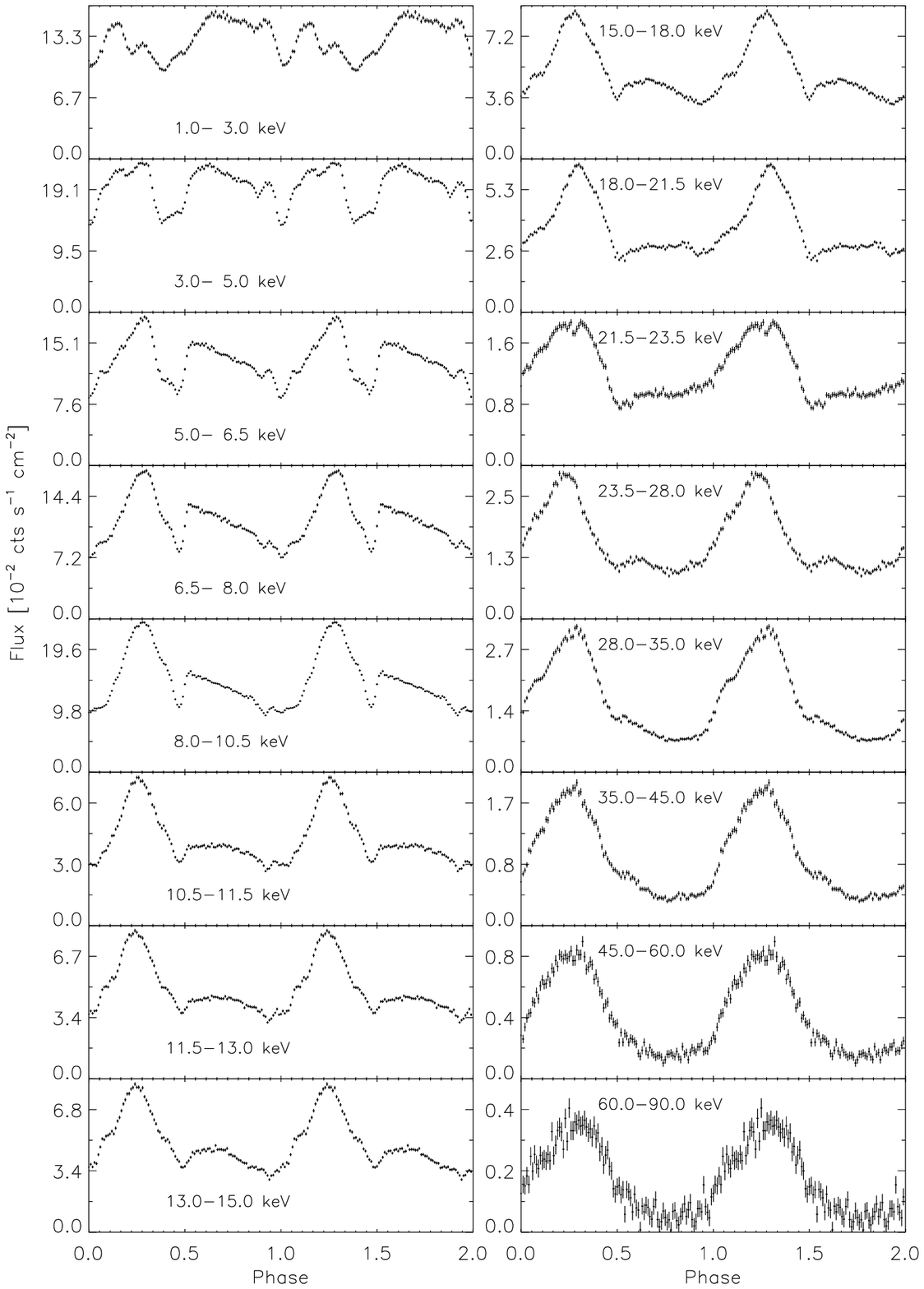}
	}
	\caption{Pulse profiles of \fu\ during observation (a) in different energy ranges. 
		Data were collected with the \mecs\ (2--8.5\,keV), the \hp\ (8.5--24\,keV) and the \pds\ (24--90\,keV).  
	         The phase-energy matrices of the three instruments were 
		arbitrarily shifted by a few energy bins (maximum 3) to take their timing systematics into account 
		and achieve an optimal alignment in their common energy ranges (8--11\,keV for the \mecs\ and 
		\hp,\ 20--25\,keV for the \hp\ and the \pds\ ).}
\label{fig:4u_pulses}
\end{center}
\end{figure}

To extract the energy dependent pulse profiles and the phase resolved spectra, 
we binned the events recorded in each energy channel of each instruments  
in 100 (\ssax) or 128 (\rxte) phase bins to form the ``phase-energy matrices''. 
These matrices were then rebinned to achieve a sufficient S/N in each energy 
bin\footnote{We optimized these values for our analysis on a trial basis. The different values 
come from the different energy resolutions of the instruments.} (at least 180 for \sax,\  and 60 for \rxte\ ). 
The exposure time of each phase bin was computed using the GTI of the corresponding observation and the reference 
time to fold the light curves was chosen so that the maximum of the high-energy pulse profiles in different observations
remains at phase $\sim 0.3$.

\subsection{Determination of phase lags.}
\label{sec:phaselag} 

For all the observation in Table~\ref{tab:observations}, we extracted a pulse profile in each of the  
energy bins of the phase-energy matrices (see Sect.~\ref{sec:timing}). 
The pulse profiles showed a remarkable dependence on the energy (especially below 
$\sim$5~keV), but did not change dramatically between the different observations. 
As an example, we show in Figure~\ref{fig:4u_pulses} the pulse profiles extracted 
for the observation (a). Above $\sim$5~keV, a prominent structure appears at phase 
$\sim$0.3 (the ``main peak'') and remains virtually stable at all the higher energies.  
The predominance of this structure is also confirmed by the analysis carried out 
in the left hand panel of Fig.~\ref{fig:pulses}. Here we represent
with red (blue) colors the phases of all 
the pulse profiles extracted from the observations in Table~\ref{tab:observations} 
where the source count rate was higher (lower).
 
\begin{figure}
  \begin{center}
  \resizebox{\hsize}{!}{
      \includegraphics[angle=0]{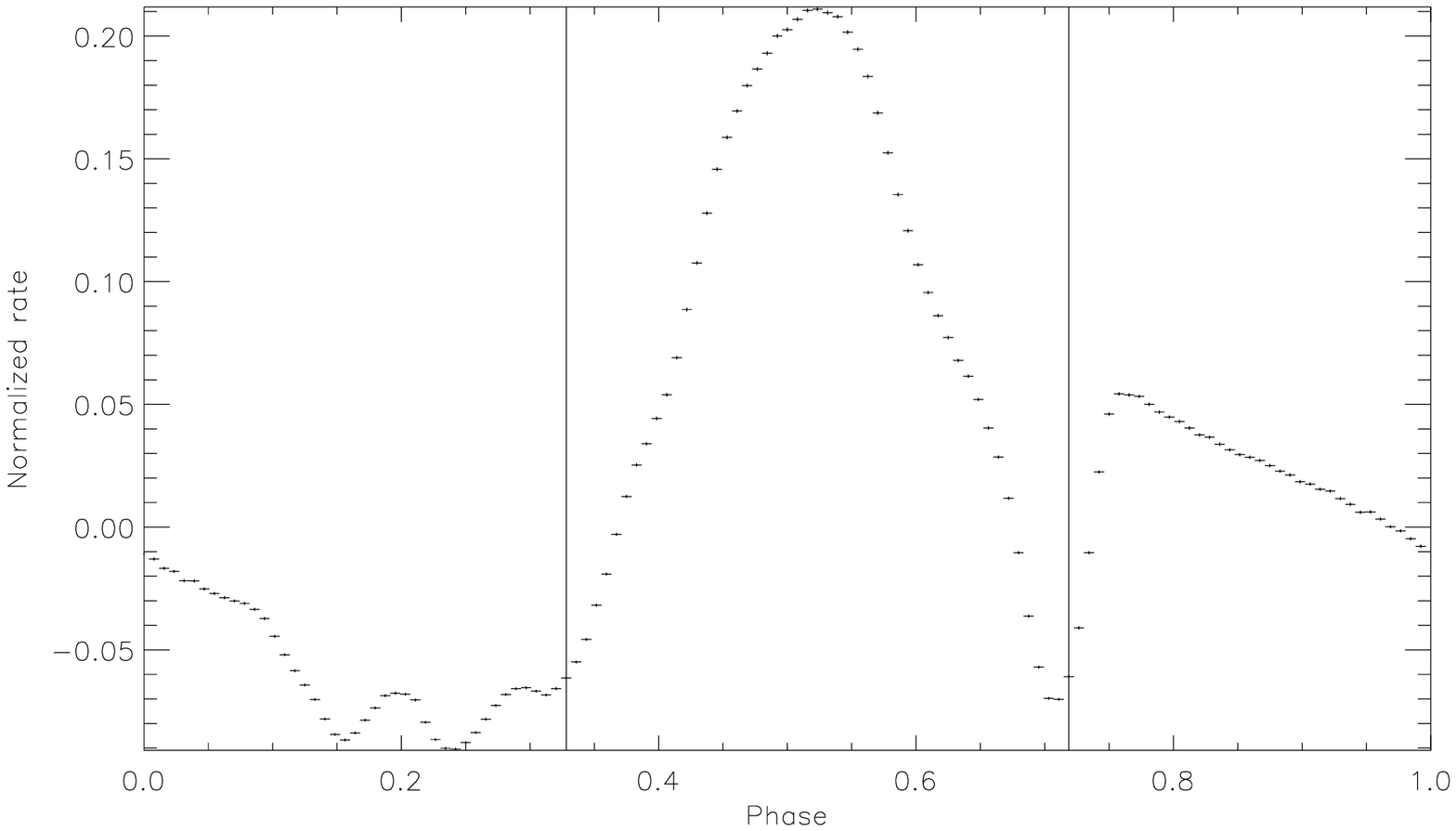}}    
	\caption{Reference pulse profile of \fu\ extracted from observation (b). 
	Data are from the \pca\ (energy range 8-10.5\,keV). The pulse profile has been linearly scaled 
	to have zero average and unitary standard deviation. The two vertical solid lines enclose the phase interval 
	used for the correlation analysis: 0.33--0.72 (the phase is shifted with respect to Fig.~\ref{fig:pulses} for clarity's sake).}
\label{fig:reference_pulse}
\end{center}
\end{figure}

\begin{figure*}
  \begin{center}
      \includegraphics[angle=0, width=0.8\columnwidth]{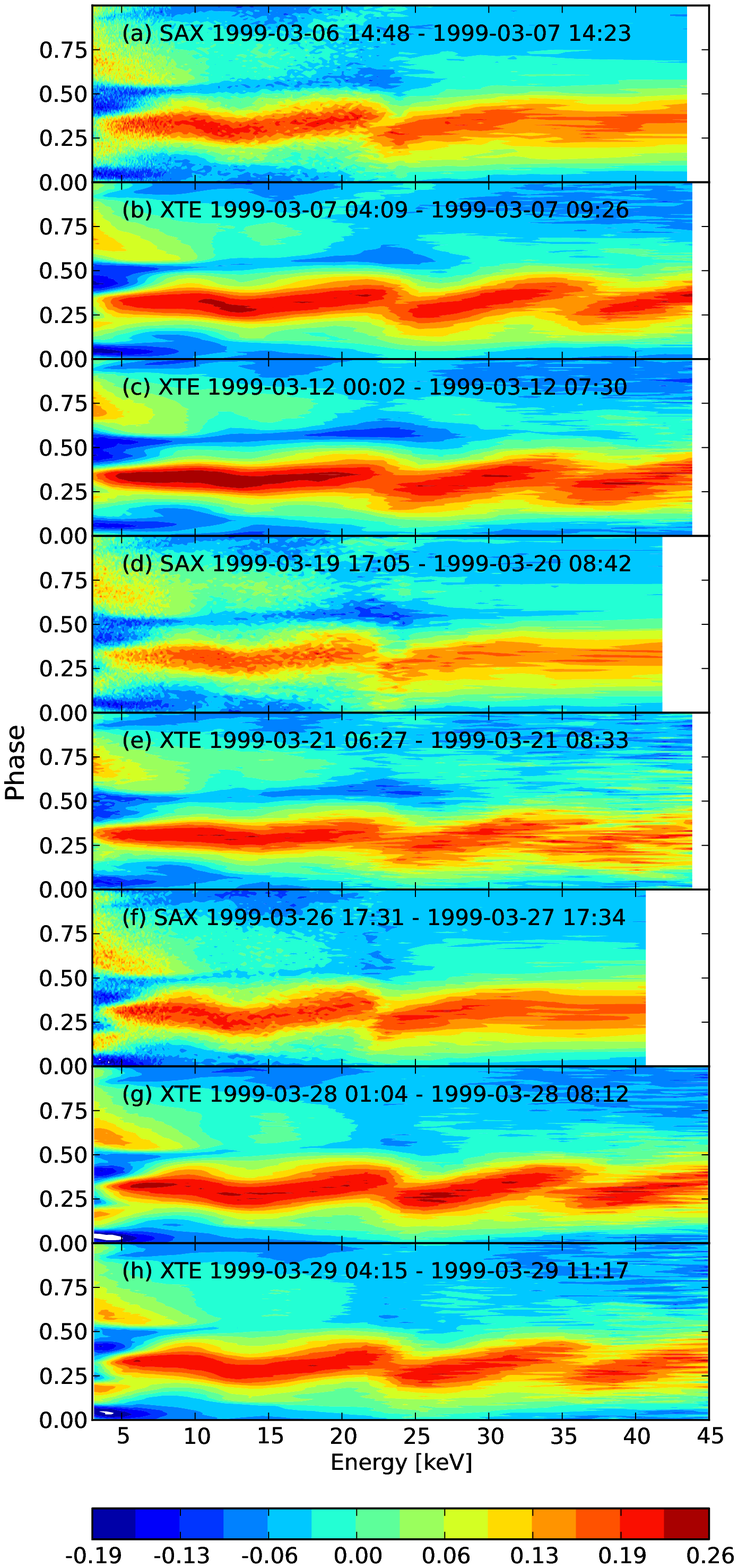}
      \includegraphics[angle=0,width=0.8\columnwidth]{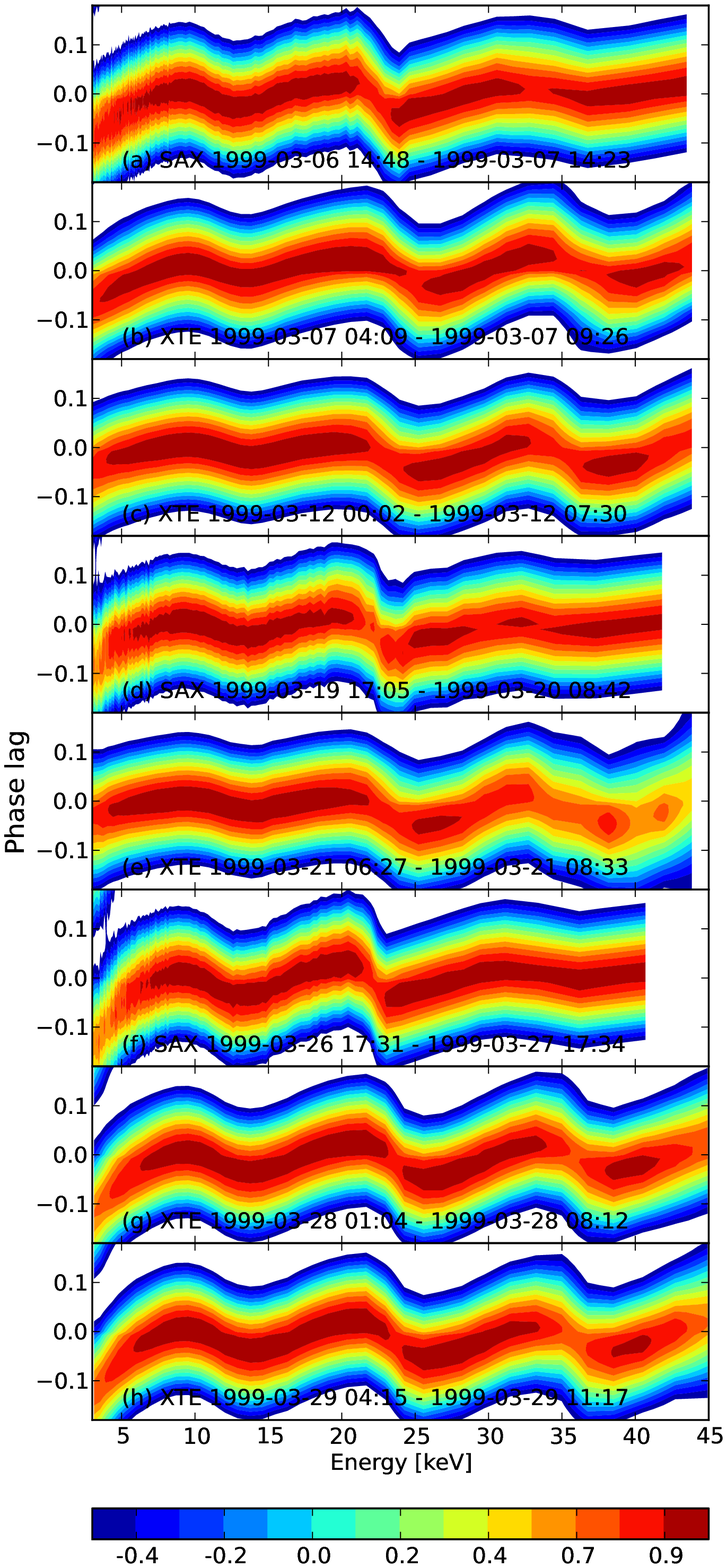}
	\caption{{\it Left panel}: pulse profiles extracted from all the observations in Table~\ref{tab:observations} 
	as a function of energy. Red (blue) colors indicate higher (lower) count rates at different 
	pulse phases and energies. As discussed in Sect.~\ref{sec:timing}, the higher intensity is around 
	phase $\sim$0.3, in correspondence with the ``main peak''. 
	The pulse profiles in this figure were all scaled to have in each energy bin a zero average value 
	and a unitary standard deviation.
   {\it Right panel}: values of the 
   correlation parameter $C_e$, as a function of phase lag with respect to the reference pulse profile and energy. 
   Red (blue) colors indicate 
   values of $C_e$ close to unity ($\ll$1). See text for details.
   }
\label{fig:pulses}
\end{center}
\end{figure*}

The relatively broad red spot, corresponding to the mean peak, 
seems to move in phase with the energy, thus showing a ``phase-lag'' in the pulse profiles. 
We note that this behavior is present in all observations. 
To investigate the origin of these phase lags in more detail, we extracted a 
reference pulse profile in the 8--10.5\,keV energy range  for each observation (see Fig.~\ref{fig:reference_pulse}) and then 
computed the cross correlation between this reference profile and the pulse profiles extracted in different 
energy bands. We limit this analysis to the phases corresponding to the main peak (see Fig.~\ref{fig:reference_pulse}) 
and use the discrete cross correlation function: 
\begin{equation}
C_e(j) = \frac{\sum_i p_r(i) p_e(i+j)}{\sqrt{\sum_i p_r^2(i)}\sqrt{\sum_i p_e^2(i)} }.     
\label{eq:correlation}
\end{equation} 
Here, $C_e(j)$ is the linear correlation coefficient calculated at the energy bin $e$ and phase bin $j$. The 
indices of the phase bin for each pulse profile are $j$ and $i$, while $p_r$ and $p_e$ are the reference and energy dependent 
pulse profiles, from which we subtracted their average values. 
The maximum value (and relative phase) of the correlation parameter for each energy bin, 
$C_{e,\mathrm{max}}$, was estimated as a function of the phase 
by performing parabolic fits. Hereafter, we use normalized phase units,
in which the phase is a real number in the range $(0,1)$.
A value of $C_{e,\mathrm{max}}<<1$ indicates that the pulse profiles $p_e$ extracted in the energy bin $e$ are 
not strongly correlated with the reference pulse profile $p_r$, 
so their shapes differ significantly. A value of $C_{e,\mathrm{max}}\sim1$ indicates that one profile can be obtained 
from the other using a linear function\footnote{The profiles in Eq.~(\ref{eq:correlation}) have zero average but not unitary standard deviation, therefore in the ideal case of $C_{e,\mathrm{max}}=1$, they are related by a function that produces a phase shift and a rescale.} and intermediate values indicate a moderately different shape.
The results of this analysis are shown in the right hand panel of Fig.~\ref{fig:pulses} in a color representation and they provide
a clear confirmation of significant phase lags in the energy-dependent pulse profiles.

\begin{figure*}
  \begin{center}
      \includegraphics[angle=0,width=\textwidth]{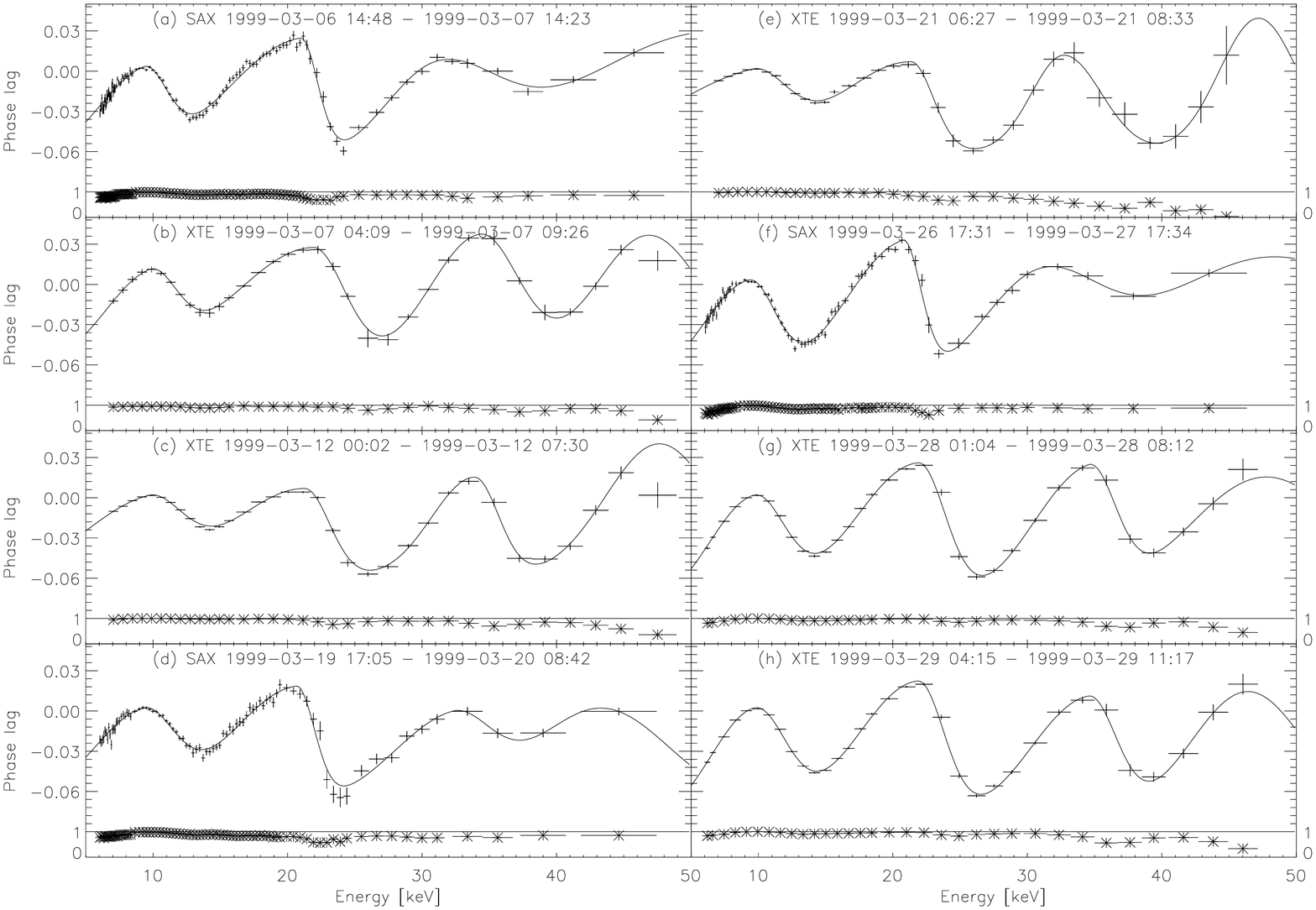}
	\caption{Phase lags of the main peak in normalized phase units (points) and linear correlation coefficients (stars) 
	as a function of energy. We considered here \sax\ data from the \mecs\ (2--8.5\,keV), the \hp\ (8.5--25\,keV), and 
	the \pds\ ($>$25\,keV). \rxte\ data are from the \pca\ (3-45\,keV) and the \hexteb ($>45$\,keV). 
	The correlation function is computed on a phase interval of 0.4 phase units, 
	centered on the maximum of the reference pulse profiles.
	The solid line represents 
	the best-fit function to the phase lags (see text for the details).} 
\label{fig:4u_lags}
\end{center}
\end{figure*}

A more quantitative representation of the measured phase lags is provided in Fig.~\ref{fig:4u_lags}. 
Here, we report the phase lags of the main peak for each observation with respect to the reference pulse, 
together with the corresponding 
uncertainties and the maximum correlation coefficient. 
To determine the phase lags and the associated uncertainties in the most appropriate way, we 
simulated 100 ``phase--energy'' matrices in which the number of events in each phase and energy bin 
was randomly selected from a  
Poisson distribution with an average value equal to the measured number of counts.
We then carried out a correlation 
analysis for each of the simulated matrices and recorded the estimated phase lags. The measurement reported for each observation 
and energy bin in Fig.~\ref{fig:4u_lags} was thus determined as the average of the simulated 
values and the corresponding uncertainty taken as one standard deviation form this value.  
We verified \emph{a posteriori} that the values measured from the real data are consistent with the 
value obtained from the average of the simulations within the reported uncertainties. 
The results reported in Fig.~\ref{fig:4u_lags} show that the pattern of the phase lags  vs. energy 
is very similar for all the observations. The main peak of the pulse is shifted toward lower 
phases at energies corresponding to the different absorption cyclotron features of \fu.\ 
In correspondence of each of these shifts in phase, 
we also notice a deviation of the cross correlation coefficient from unity. 
This testifies that, at energies close to those of the different cyclotron features, the shape of the 
main peak is slightly different from that of the reference pulse profile. 
At the highest energies ($\gtrsim$40~keV), the marked decrease in the correlation coefficient 
is due to statistical noise.
To correctly evidence the features visible that are in the pattern of the phase lags as a function of energy, we fit 
a model comprising several asymmetric Gaussian lines  to these data\footnote{The empirical function used to fit the phase-lag energy dependency is
$f(e)=A+\sum_{i=1}^{4}N_i \exp \left[ -\left( e - E_i \right) ^2 / \sigma_{i,\lbrace d, u \rbrace }^2\right] $,
where $\sigma_{ i,\lbrace d, u \rbrace }=\sigma_{i,d}$ for $e \leq E_i$,  
$\sigma_{ i,\lbrace d, u \rbrace }=\sigma_{i,u}$ for $e > E_i$, and $e$ is the energy. 
The fourth Gaussian is only poorly determined due to the low S/N of the data.}.  

We note that the intrinsic energy resolutions of the different instruments 
(not accounted for in our phenomenological fit of the phase lags) affects the 
estimate of the centroid energies of the Gaussian empirical functions. From Fig.~\ref{fig:4u_lags}, it is clear 
that the third Gaussian is systematically detected with a centroid energy lower in the \pds\ than in the \pca\ 
data, because the former instrument is characterized by a lower energy resolution than the latter. The opposite effect is visible on the second Gaussian
by comparing the \pca\ with the \hp\ data, as the latter has a finer energy resolution than the former.  
Taking these systematic effects into account, we conclude that the phase lags measured from the different instruments 
and observations in Table~\ref{tab:observations} are in good agreement with each other, and no prominent 
variations of the phase lag pattern can be appreciated in the luminosity range of \fu\ spanned by the data. 

\begin{figure}
  \begin{center}
\resizebox{\hsize}{!}{\includegraphics[angle=0]{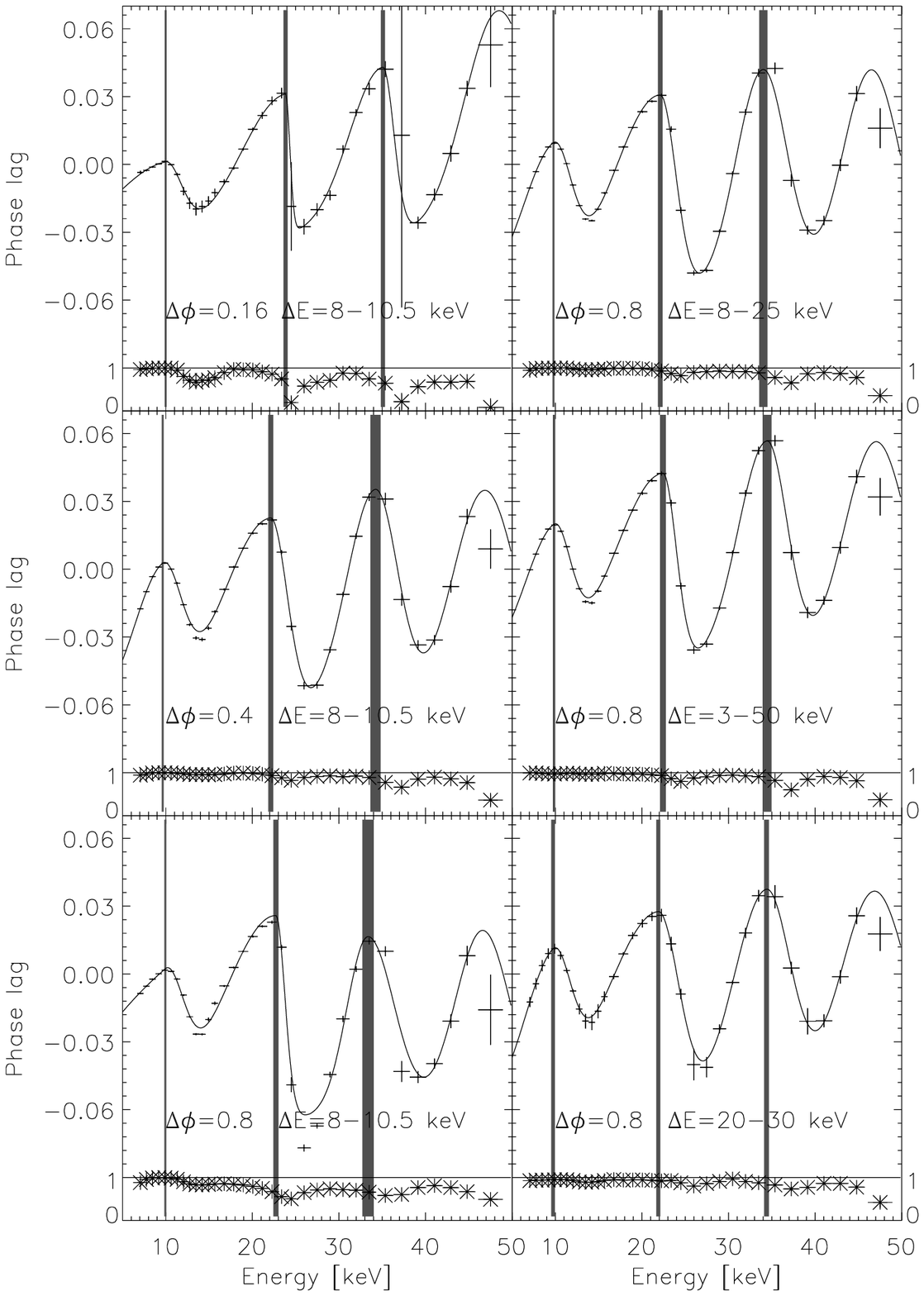}}
         \caption{Phase shifts measured from the observation (b) by using a different  
         range of phases than adopted for the analysis in Fig.~\ref{fig:4u_lags} 
         to compute the shifts ($\Delta\phi$) and a different energy range of the reference 
         pulse profile ($\Delta E$). Specific values of these parameters 
         are indicated in different panels.  The shaded 
         intervals represent the centroid energy of the asymmetric Gaussians used to describe 
         the lags with the $1\sigma$ uncertainty. The solid line is the best-fit function of 
         the phase lags.}
\label{fig:check_w_e}
\end{center}
\end{figure}

Finally, we checked that all the above results do not depend on the particular settings adopted in
the method developed for the analysis described above. 
For this purpose, we performed a further analysis of the observation 
(b) by adopting different values of the range of both the phases used for the correlation analysis 
($\Delta\phi$) and the energy range chosen to extract the reference pulse profile ($\Delta E$). 
All the results of this analysis are reported in Fig.~\ref{fig:check_w_e}. 
We note that the adoption of a smaller $\Delta\phi$ resulted in a 
significant decrease in the reliability of our analysis, as a large part of the peak falls 
outside the sampled window (also indicated by the lower value of the 
cross-correlation coefficient). Choosing a larger $\Delta\phi$ 
would instead lead to a loss of coherence in the cross-correlation due to the complex 
energy dependency of the pulse profiles outside the main peak. 
Changing the value of $\Delta E$ resulted, as expected, only in a systematic offset of the phase shifts. 
We thus concluded that the general trend of the phase shifts 
is not affected by the particular set of parameters used for the analysis.  

\subsection{Spectral analysis}
\label{sec:spectroscopy} 

In order to deeply investigate the relation between 
the phase lags and the energy of the cyclotron scattering 
features, we also performed a spectral analysis of the source. 

\begin{figure}
  \begin{center}
  \resizebox{\hsize}{!}{
      \includegraphics[angle=0]{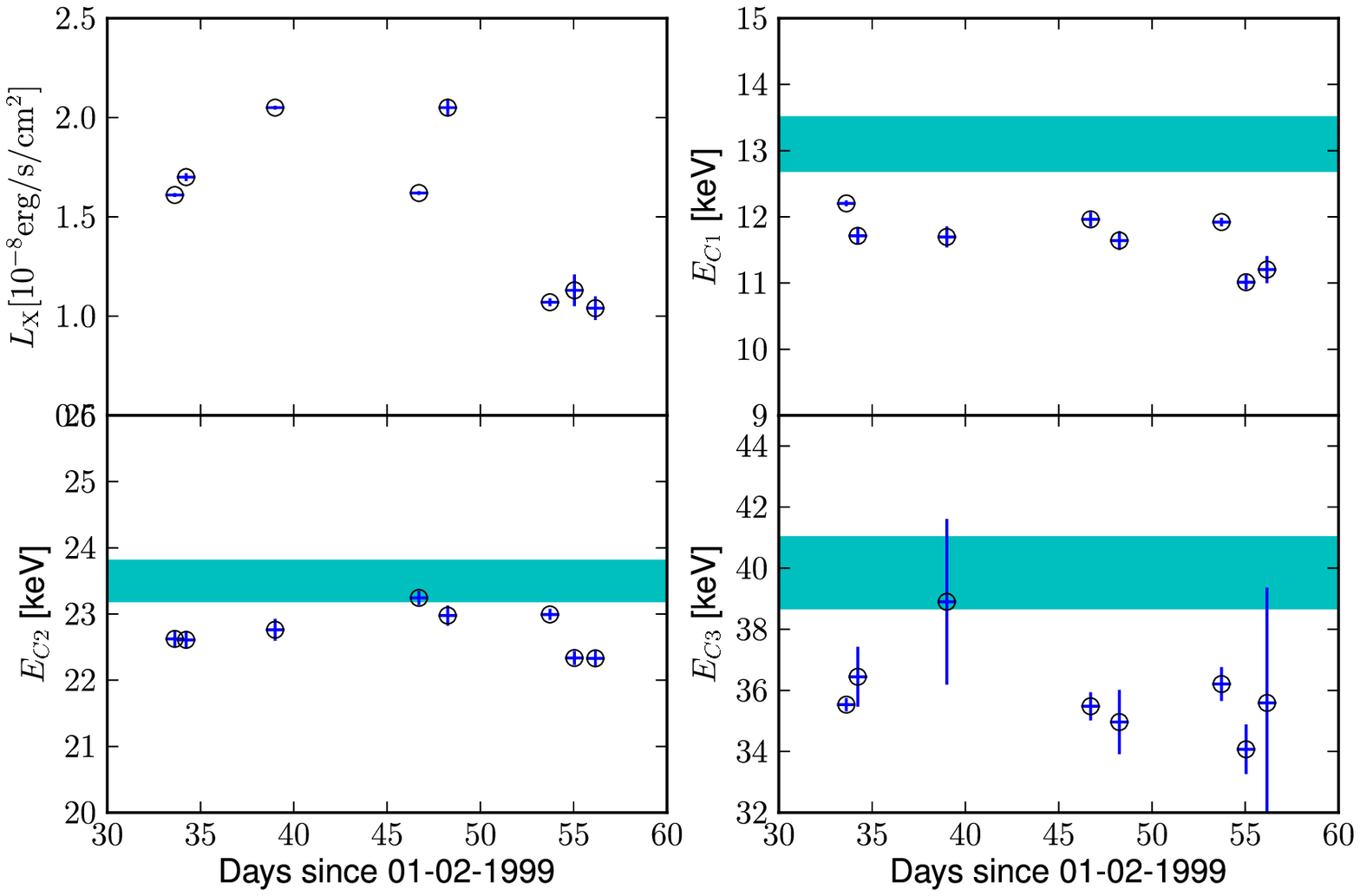}}
\caption{Phase-averaged luminosity and centroid energy of the absorption lines. The horizontal cyan bands are the energy 
range of the most negative phase shift of the main peak measured from the \sax\ data. Uncertainties on fitted parameters 
are at 90\% c.l. \emph{Upper left}: X-ray luminosity (see Table~\ref{tab:observations}). \emph{Upper right}: centroid energy of the fundamental. 
\emph{Lower left}: centroid energy of the first harmonic. \emph{Lower right}:centroid energy of the second harmonic.}
\label{fig:all_lines}
\end{center}
\end{figure}

We first extracted 
the phase-averaged spectra\footnote{For the phase-averaged spectral analysis of the \rxte\ 
observations, we used the \texttt{standard2} mode of the \pca\ data and both \hexte\ units. 
For the spectral fitting, we used the \xspec\ software (v.12.6.0u)}for each observation in Table~\ref{tab:observations} and 
performed a fit to these data with the phenomenological multicomponent model proposed for \fu\ by  
\citet{ferrigno2009}. For \pca,\ we added a 0.5\% 
systematic error and fixed the absorption column at 
$N_\mathrm{H}=9\times10^{21}\,\mathrm{cm^{-2}}$, the best-fit value obtained 
in the \sax\ observations. This provided, in each case, an accurate measurements of the flux (and thus the luminosity) 
of the source (see Table~\ref{tab:observations}). 
For each observation, we also report the centroid energy of the first three 
absorption lines in Fig.~\ref{fig:all_lines}, together with the energy range at which the most negative phase lag 
occurs in the corresponding energy range. Although the most negative phase shifts occur slightly above the measured centroid energies for all lines, there is a clear indication of the link between timing and spectral features. In the following paragraphs, we clarify the origin of this apparent mismatch of energies.

A phase-resolved spectral analysis was carried out to study the variation 
in the properties of the cyclotron absorption lines at different phases 
\citep[see e.g. ][]{mihara2004}. Following the discussion in Sect.~\ref{sec:timing}, 
only the range of phases including the main peak was considered.  
We extracted source spectra in 64 and 50 
phase bins for the \pca\ and the \hp,\ respectively. 
We restricted the energy range for the phase-resolved analysis
to the 10--35\,keV interval and adopted in the fit a simplified 
spectral model comprising a power-law modified by a high-energy cut-off and three Gaussian 
absorption lines (\texttt{gabs} in \xspec). This permitted us to clearly identify the 
centroid energy of the fundamental and of the first harmonic in most spectra. 
We did check \emph{a posteriori} that the simplified spectral model
did not change our results significantly with respect to those already reported, e.g.,  by \citet{ferrigno2009}.

\begin{figure}
  \begin{center}
  \resizebox{\hsize}{!}{
      \includegraphics[angle=0]{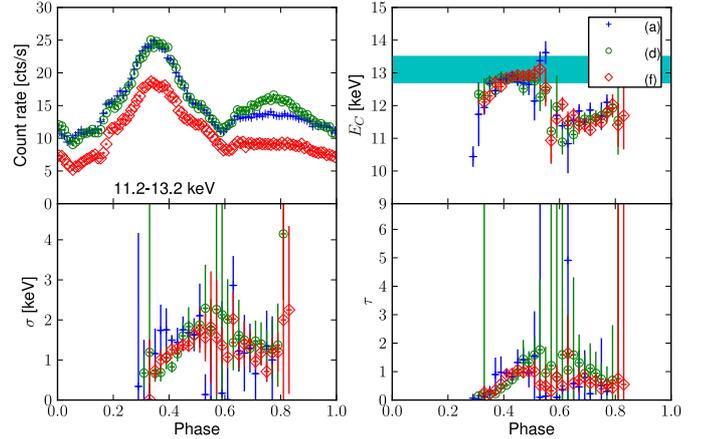}}
	\caption{Phase-resolved study of the fundamental absorption line 
	for the \sax observations. Blue points correspond to observation (a), 
	green circles to observation (d), and red diamonds to observation (f). 
	\emph{Upper left}: Pulse profiles extracted in the energy range 11.2-13.2\,keV (shifted in phase to 
	be aligned). \emph{Upper right}: Centroid energy of the first cyclotron harmonic. The horizontal 
	cyan band is the energy range of the most negative phase shift of the main peak measured from the \ssax\  
	data. \emph{Lower left}: Amplitude of the line. \emph{Lower right}: Optical depth 
	of the line.}
\label{fig:sax_line_1}
\end{center}
\end{figure}

\begin{figure}
  \begin{center}
  \resizebox{\hsize}{!}{
      \includegraphics[angle=0]{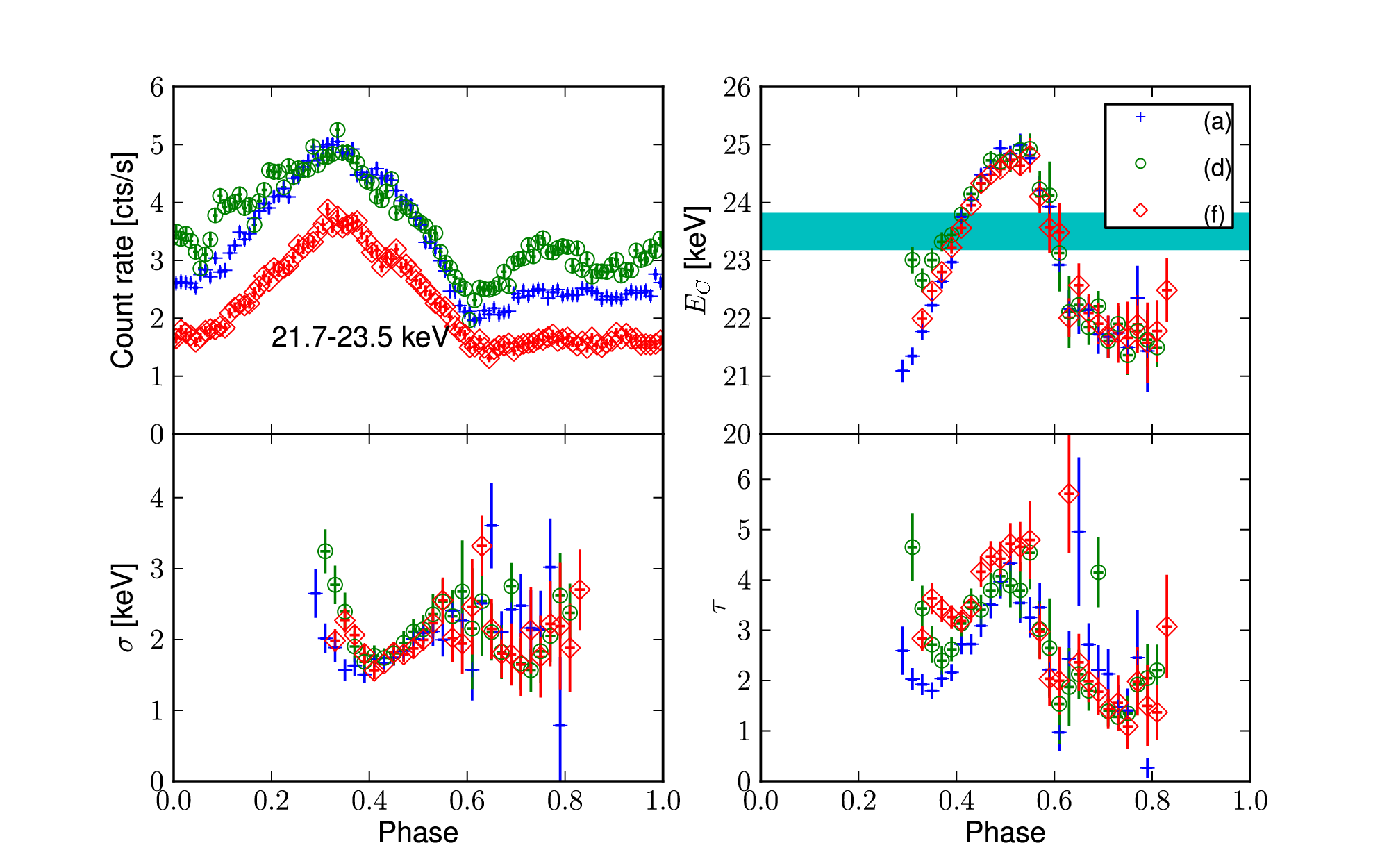}}
\caption{Phase-resolved study of the first harmonic for the \sax observations. Blue points correspond to observation (a), 
green circles to observation (d), and red diamonds to observation (f). Uncertainties on fitted parameters 
are at 90\% c.l. \emph{Upper left}: pulse profiles extracted by using data from the \hp\ in the energy range 
21.7-23.5\,keV. \emph{Upper right}: centroid energy of the first cyclotron harmonics. The horizontal cyan band is the energy 
range of the most negative phase shift of the main peak measured from the \sax\ data. 
\emph{Lower left}: amplitude of the line. \emph{Lower right}: Optical depth of the line.}
\label{fig:sax_line}
\end{center}
\end{figure}

In Fig.~\ref{fig:sax_line_1}, we show the values of the centroid, amplitude ($\sigma$), and optical 
depth ($\tau$) of the fundamental vs. the pulse phase, while
in Fig.~\ref{fig:sax_line}, we show the corresponding 
results for the first harmonic for the three \sax\ observations. 
The pulse profiles in the relative energy ranges are also shown. 
For the second harmonic, we verified 
that the poorer S/N prevents such detailed analysis, by also using the \pds data to cover the range above 34\,keV.  
The parameters of the lines could not be well constrained at different phases from those of the 
main peak, as in these cases the lines appeared absent or shallower and less pronounced. 
Despite the slightly different shape 
of the pulse profiles and the variation in the luminosity of the source (see Table~\ref{tab:observations}), 
the three observations gave fairly similar results for the line parameters. This confirmed 
that the scattering 
region is located at roughly the same height above the NS in this luminosity range.
In the upper right hand panel of these figures, we use a horizontal band to indicate
the energy range in which the largest negative shift in phase of the pulse profiles 
in the \sax\ observations was measured. 
In both cases, the centroid energies correspond to the 
energy of the most negative phase shift in some phase interval. For the fundamental, such a phase range corresponds 
to the highest optical depth of the line, but to an intermediate value for the first harmonic. 
We notice that the phase resolved spectroscopy proves how the phase-averaged spectral results 
are a non-trivial average of phase variability and cannot be directly compared to our results for phase lags. 
For all the lines, the centroid energy estimated from the average spectra is lower than the energy of 
the corresponding phase lags, but the results of the phase-resolved spectroscopy clearly indicate
that the energy dependency of the shift in phase of the pulse profiles is  
related to the cyclotron-scattering process. 
Similar results were obtained by performing a phase-resolved analysis of the \rxte\ data over a slightly wider range of
luminosity.

It is particularly interesting to note that in all cases the cyclotron lines get   
narrower and deeper in the descending edge of the pulse (i.e., the values of $\sigma$ and 
$\tau$ are respectively lower and higher in these phases). We discuss this observational property of 
the line at the end of Sect.~\ref{sec:model}.

\section{A model for the emission from an accretion column}
\label{sec:model}

In this section, we describe the model we developed to interpret the energy-dependent phase shifts of the pulse profiles 
from \fu\ and the properties of its cyclotron emission lines (see Sect.~\ref{sec:phaselag} 
and \ref{sec:spectroscopy}) in terms of an energy-dependent beamed emission emerging 
from two accretion columns on the NS. The model has a two-step structure: in the first
part, we compute the flux emerging from the column as a function of its inclination angle with the 
line of sight (the ``asymptotic beam pattern''). In the second part, we exploit the geometric parameters
of \fu\ derived by \citet{sasaki2011} using the pulse decomposition method to model the observed pulse profiles.

\subsection{Asymptotic beam pattern of an accretion column.}
We considered in the model a spherical NS, with radius $R_\mathrm{NS}$ and mass $M_\mathrm{NS}$. 
Based on the conclusion of \citet{sasaki2011}, we exclude the ``hollow cone'' geometry and then assume that 
the accretion column is homogeneously filled 
by the accreting material and characterized by a conical shape. 
The latter is 
a good approximation in case the accretion flow is confined by a dipolar magnetic field 
\citep{leahy2003}. Finally, we assume that the radiation is emitted from the 
lateral surface of the column, as expected for the X-ray luminosity of \fu\ 
in the observations analyzed here \citep[$L_X \simeq 10^{38}\,\mathrm{erg\,s^{-1}} >  4\times 10^{37} \,\mathrm{erg\,s^{-1}} $][]{basko1976}. 
As the emission from the accretion column is relatively close (at most a few km) to the NS surface, 
we also take the most relevant relativistic effects into account. We include in 
our model the relativistic beaming, the gravitational redshift and the gravitational lensing effect (also known as solid 
angle amplification), according to the method described by \citet{leahy2003}. However, for the light bending, 
we use the approximate geodesic equation proposed by \citet{beloborodov2002}\footnote{The range of validity of this approximation was 
extended as described in Sect.~\ref{sec:appendix} to account for the photons emitted above the 
NS surface in a downwards direction.}. 
The mathematical details of the models are reported in 
Appendix~\ref{sec:appendix}. 
The model parameters are thus the NS mass and radius, the semi-aperture of the accretion column, 
$\alpha_0$, the column height with respect to the NS surface, and the emissivity of the 
surface element of the column in its rest frame, $I_{0,\nu}(r,\phi_0,\alpha,\phi_1)$ 
(i.e., the intrinsic beam pattern). Here,  
$\nu$ is the radiation frequency, $r$ the height at which the photon is emitted, $\phi_0$ 
its azimuth around the column axis, $\alpha$ the angle between the local radial direction and the photon 
trajectory, and $\phi_1$ is the corresponding azimuth (see Fig.~\ref{fig:psi_alpha}). 

Deriving $I_{0,\nu}(r,\phi_0,\alpha,\phi_1)$ on a physical basis would require a detailed 
treatment of all the relevant photon-scattering processes in the accretion column, which is beyond 
the scope of the present paper. 
Instead, we consider a phenomenological model of $I_{0,\nu}(r,\phi_0,\alpha,\phi_1)$ depending just on the angle $\alpha$.
The dependency on $\phi_1$ can be neglected as the cyclotron scattering in the geometry considered here is almost independent from this angle, because the magnetic field lines are nearly perpendicular to the NS surface 
at the polar caps. We also assume that the intensity of the radiation emitted from the column 
does not depend on the radius $r$ and the angle $\phi_0$. Including the dependence on $\phi_0$ 
could be relevant for tracing the coupling between the accretion disk with the magnetosphere. The dependency on $r$ would 
account for the vertical density and velocity profiles in the accretion column \citep[see also][]{leahy2004a,leahy2004b}. 
We have verified that
introducing a dependency of the emerging beam intensity on $r$ does not have significant implications for 
the problem of modeling the observed phase shifts.

Our model of the radiation beaming comprises two emission components: one directed downwards and 
one upwards. Such beaming, at the end of our calculations, 
qualitatively reproduces the asymptotic beam patterns obtained by \citet{sasaki2011} 
with the decomposition method (see Eq.~\ref{eq:ialpha}). 
The actual beaming is determined by the non-trivial interplay of special relativistic 
transformations due to the bulk motion of the emitting flow
and anisotropic scattering. To estimate the magnitude of the special relativistic effects, we adopt the column
model of \citet{becker1998} for two values of their parameter $\epsilon_c$ (0.5 and 1) defined in their Eq~(4.1). At the top of a 
2\,km column, the radiation advected downwards is two to three times 
more intense than the radiation escaping upwards for an isotropic field in the matter rest frame.
The escaped radiation becomes nearly isotropic at the bottom of the column, where the flow settles. 
According to this model, 
most of the radiation is emitted in the lower part of the column, below the radiative shock, which is predicted to reside 
at a few hundred meters above the NS surface. Weighting the downward and upward emission with their emissivity 
does not affect our main assumption, as we find that the total downward radiation is 40--70\% more intense than the upward one.

The bulk motion of the emitting plasma is not the only driver of the radiation beaming, because the scattering cross section 
are highly asymmetric in the strong magnetic field of HMXBs. Several studies have predicted a certain degree of anisotropy \citep[see][and references therein]{kraus2001} for the radiation emitted far from the cyclotron resonances. More recently, \citet{gabi2007} have carried out MonteCarlo simulations and showed
that the degree of anisotropy of the radiation in the energy range affected by the lines
also depends on the optical depth of the plasma and does not exceed
several percent (see their Fig.~7). No detailed studies have been carried out,  to our knowledge,
on the angular beaming of the radiation at energy close to the ones of the cyclotron scattering features, with the resolution 
required by the modern instruments. For this reason,
our qualitative description of the radiation beam through the introduction of upwards and downwards components
was mainly inspired by the study of the radiation spectral properties as a function of the viewing angle
\citep[see][and references therein]{gabi2007}.
For the results of the existing modeling\footnote{We concentrate our attention to the first harmonic, 
for which there is no photon spawning, which would add an unnecessary complication to our reasoning.}, 
we notice that the absorption line is deeper in the direction perpendicular to the
magnetic field, thus resulting in an effective suppression of photon number in that direction. 
These photons are scattered primarly
along the direction of the magnetic field, where the cross section is suppressed, and would presumably lead to an enhancement
of the beamed radiation along the column axis. As the Doppler boosting is not extremely severe in the velocity field of the 
infalling plasma, we conclude that a considerable fraction of the photons will be able to escape in the vertical direction
and originate to an upward directed component.

To summarize, we associate the downward component with the radiation advected by the plasma flow 
\citep[as already proposed, see e.g.,][]{kraus2001} and the upward component, introduced here for the first time,
with the (cyclotron-) scattered radiation. The latter component should be prominent close to the NS surface, where the flow 
is characterized by a reduced bulk motion, but gives rise to a non-negligible contribution also along 
the column extension, owing to the relatively 
moderate Doppler boosting.
We model the first component with a Gaussian centered on $\alpha=150\gra$ and characterized by a width 
of $45\gra$ (normalization set arbitrarily to 100), and the second component with a Gaussian centered 
on the local zenith and characterized by a width of $45\gra$ and a variable normalization.  
Even though we use this heuristic approach, our model is able to qualitatively account for
the most relevant properties of the intrinsic beam pattern emerging from the accretion column 
of \fu.\  

Once $I_{0,\nu}(\alpha)$ is given, the asymptotic beam pattern can be computed 
as a function of the angle between the axis of the column and the line of sight to the observer 
($\psi_\mathrm{obs}$, see Appendix~\ref{sec:appendix} for the details). 
In Fig.~\ref{fig:beams}, we show the intrinsic beam patterns in the left hand panels, 
$I_{0,\nu}(\alpha)$,
for different values of the normalization of the upward component. 
The corresponding asymptotic beam patterns are reported in the central panels. 
These were computed by assuming an accretion column of semi-aperture $4\gra$, extended  
for 2\,km above the surface of a canonical NS with mass $1.4\,\msol$ and radius 10\,km 
\citep[column height from][]{nakajima2006}. We notice that the shape 
of the beam is not very sensitive to the different parameters 
adopted for the column and the NS, provided that the 
upper part of the accretion column remains visible at large viewing 
angles ($\psi_\mathrm{obs} \ga 150\gra$). 

\subsection{Simulated pulse profiles.}
\begin{figure}
  \begin{center}
\resizebox{\hsize}{!}{\includegraphics[angle=0]{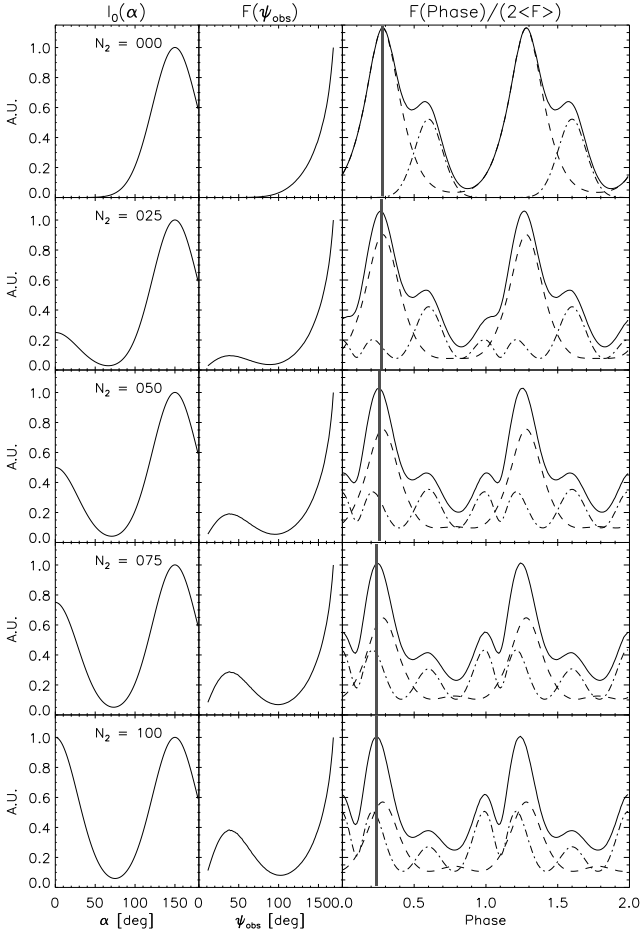}}
 \caption{\emph{Left panels}:  intrinsic beam patterns (arbitrary units) emerging from the 
lateral walls of the NS accretion column as a function of the angle $\alpha$ with respect to 
the local radial direction. The function is the sum of two Gaussians as described 
in Eq.~(\ref{eq:ialpha}). The value of the normalization of the upward Gaussian is indicated 
in each plot ($N_2$). \emph{Central panels}: asymptotic beam patterns (arbitrary units normalized 
to the values at 170~degrees) as a function of the angle $\psi_\mathrm{obs}$ 
between the axis of the accretion column and the line of sight to the observer. The beams are computed by 
assuming an accretion column with a semi-aperture of 4~degrees, and extended for 2\,km above 
the surface of a neutron star with a mass of $1.4\,\msol$ and a radius of 10\,km.  
 \emph{Right panels}:  synthetic pulse profiles computed for the geometry in \fu\ described in Sect.~\ref{sec:model}. 
 The solid line corresponds to the total flux divided by two times its averaged value for plotting purposes. The dashed line 
 represents the contribution from the farther pole from the line of sight to the observer, while the dot-dashed line 
 indicates the contribution from the other pole. The gray vertical band corresponds to the location in phase 
 of the pulse maximum. A leftward shift in phase is emerging while increasing the value of $N_2$.}
\label{fig:beams}
\end{center}
\end{figure}

Assuming the above asymptotic beam patterns, we computed the pulse profiles 
of \fu\ by using the geometrical parameters of the system described in Figs.~2 and 
3 of \citet[][all measured with respect to the rotation axis of the NS]{kraus1995}.  
These are the direction of the line of sight to the source ($\Theta_0$), the colatitude of 
the two magnetic poles ($\Theta_1$, $\Theta_2$), and their difference in longitude ($\Delta$). For the 
accretion column located on the $i$-magnetic pole, we can write the following relation between these angles: 
\begin{eqnarray}
\lefteqn{\cos \psi_\mathrm{obs} = \cos \Theta_0 \cos \Theta_\mathrm{i} +}  \nonumber \\
& \sin \Theta_0 \sin \Theta_i \cos \left[ \Omega (t-t_\mathrm{ref}) -(i-1)(\pi - \Delta ) \right]\,,
\label{eq:psi_obs}
\end{eqnarray}
where $i=(1,2)$, and $\Omega=2\pi/P_\mathrm{spin}$ is the pulsar angular velocity measured at a reference time $t$=$t_\mathrm{ref}$.\footnote{The spin period of accreting pulsars is not constant due to the torque exercised by the in-falling mass.}
We assume in the following $\Theta_0=60\gra$, $\Theta_1=148\gra$, $\Theta_2=74\gra$, and 
$\Delta=68\gra$ \citep{sasaki2011}. As an example, we show 
the pulse profiles computed in the right hand panels of Fig.~\ref{fig:beams} by using the above  
geometrical parameters and the beam patterns reported in the central panels of the same figure. 
We did not consider other geometrical configurations, because we constructed the intrinsic beam pattern to 
reproduce the asymptotic beam pattern obtained for $\Theta_0=60\gra$. However, we verified that phase lags
can be modeled also for other allowed configurations by adopting a different intrinsic beaming.

By comparing Figs.~\ref{fig:4u_pulses} and \ref{fig:beams}, we note that 
the model developed above can reproduce the shape of the pulse profiles 
of \fu\ relatively well at energies higher than the fundamental cyclotron scattering feature. In our model, 
the bump in the profile that follows the main peak is due to the emission 
from the accretion column closer to the observer's line of sight. 
However, we note that the double-peak structure of the upward emission (represented by the dot-dashed line at phase 
0.9-1.2 in Fig.~\ref{fig:beams}), which results from the model, does not seem to be 
present in the observed pulse profiles. The suppression of the flux 
between these two peaks (at phase $\sim1.1$) is unavoidable in the model, because it comes from the decreased 
flux of the asymptotic beam pattern at angles $\psi_{\rm obs} \lesssim 30\gra$ produced by 
the self-obscuration of the column. The double peak could be avoided if the upward 
beam were emitted slightly outside the column (e.g., in a scattering halo near or on the NS surface), or if part of the radiation 
also escapes along the accretion column axis in a pencil beam.
Our simplified model could not reproduce this kind of emission, as this would have required either a termination of the column 
at some height with emission from its top or the introduction of an emitting region around the column. However, we
note that our calculation
was not aimed at reproducing the exact shape of pulse profile, but rather at finding a plausible 
explanation for the phase lags observed in correspondence of the cyclotron lines, an effect which has never been reported before.

Indeed, the phase lags measured from the data are also reproduced relatively well in the model 
(see Fig.~\ref{fig:beams}).   
According to our calculations, the displacement of the main peak in phase is due to the 
distortion of its shape at energies close to those of the  
cyclotron absorption lines, where the contribution of the upward Gaussian component 
to the intrinsic beam is enhanced. In our model, the larger estimated shift 
in phase is -0.05 (measured in phase units, for $N_2=75$-100), and the corresponding value of the 
cross-correlation coefficient is 0.85. The analysis was limited here  
to the phases 0.0-0.5 that correspond to the mean peak of the pulse profiles, 
see right panels of Fig.~\ref{fig:beams}. 
This is in fairly good agreement with the results found from the data.
We also note that the location of the main pulse peak moves to lower phases at low energy ($E\lesssim5$\,keV),
where a scattering halo on the NS surface dominates. This is a further indication that the scattered radiation is responsible for the observed phase shifts.

\begin{figure}
  \begin{center}
\resizebox{\hsize}{!}{\includegraphics[angle=0]{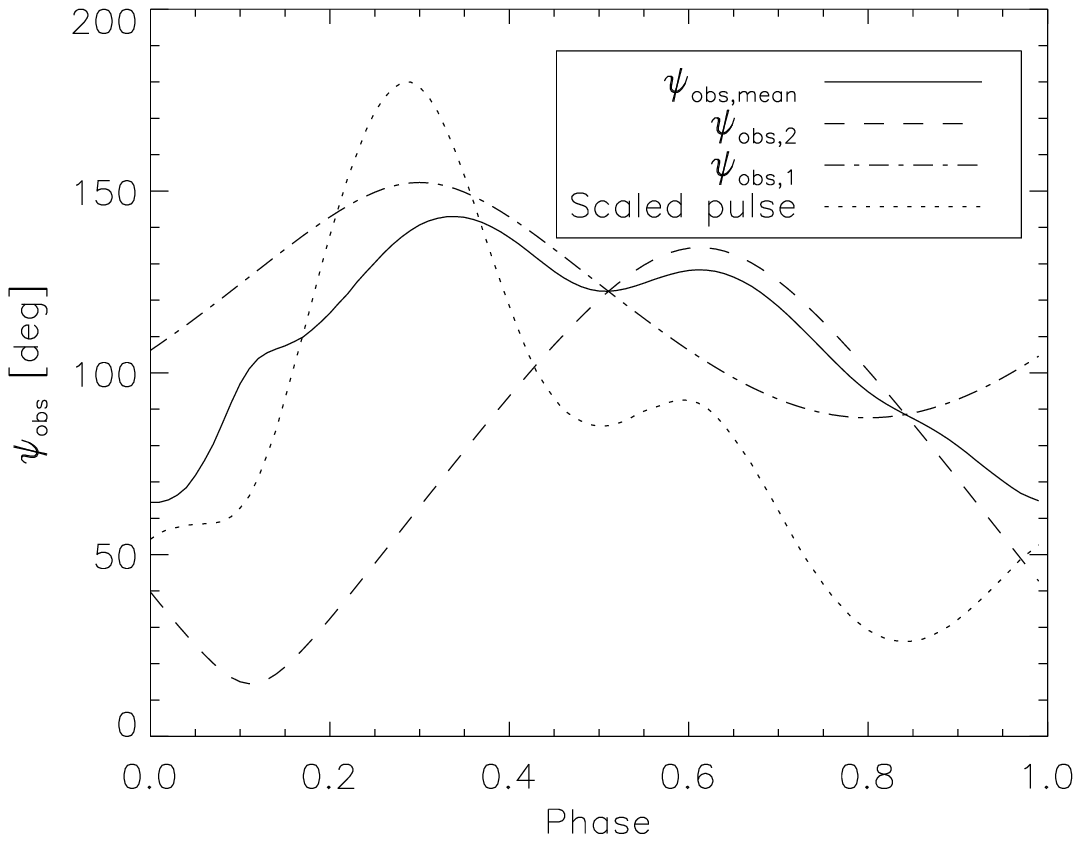}}\\
\resizebox{\hsize}{!}{\includegraphics[angle=0]{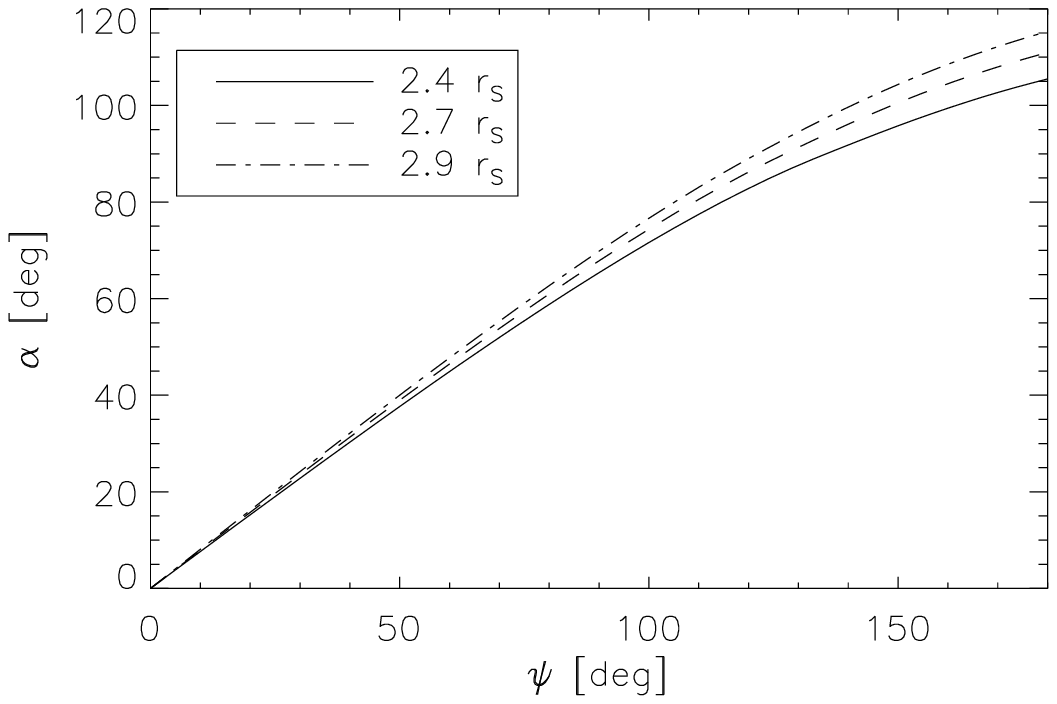}}
         \caption{\emph{Upper panel}: inclination angles of the columns with respect 
         to the line of sight to the observer obtained by assuming an intrinsic beam 
         pattern with $N_2=25$. The dashed line 
         represents the column that is farther away from the observer; the dot-dashed 
         line, the other. The solid line is the average inclination angle of the two 
         columns weighted for their relative flux measured at the location of the observer. 
         The dotted line is the synthetic pulse profile, 
         scaled to fit into the plot. 
         \emph{Lower panel}: relation between the angles 
         $\psi$ and $\alpha$ of Fig.~\ref{fig:psi_alpha} for different locations from 
         the center of the NS in units of the gravitational radius. The three locations 
         correspond to the base of the accretion column, its midpoint, and the top of the column, 
         respectively (calculated according to the geometry of the system defined 
         in Sect.~\ref{sec:model}).}
\label{fig:psi_pulse}
\end{center}
\end{figure}

Finally, we show that our model is compatible 
with the phase-resolved spectral properties of the absorption lines carried out in 
Sect.~\ref{sec:spectroscopy}. 
In Fig.~\ref{fig:psi_pulse} (upper panel), we report the viewing angles ($\psi_{\mathrm{obs, }i}$)
of the two accretion columns onto the NS with respect to the line of sight as 
a function of the phase. We also plot the average viewing angle (solid line) calculated for 
the combined radiation of the two columns (weighted for their 
relative contribution estimated by using the beam pattern in our model), 
which results
in a relatively large angle ($\sim120-150\gra$) on the descending portion of the main peak 
(phases $\sim$0.3-0.6), and significantly lower values at other phases.  
To understand the meaning of these considerations, 
in the bottom panel of Fig.~\ref{fig:psi_pulse}, we plot the relation between the angles  
$\alpha$ and $\psi$, which are calculated from the relativistic ray-tracing of the model 
(see also Fig.~\ref{fig:psi_alpha}).  
The former angle represents a good approximation of the relative direction between the radiation 
emerging from the column and the local magnetic field, while $\psi$ is the trajectory deflection angle
(see Fig.~\ref{fig:psi_alpha}), which differs from the column viewing angle ($\psi_\mathrm{obs}$) by at most a few degrees. 
We note that $\psi\sim120-150\gra$ corresponds to $\alpha\sim80-100\gra$; therefore
at the phases of the right flank of the main peak, the 
emerging radiation is emitted in a direction nearly perpendicular to 
the local magnetic field. At these inclinations,  
the resonant scattering features are expected to be deeper and narrower \citep[see ][and references 
therein]{gabi2007}, in agreement with the results found from the phase-resolved spectral 
analysis carried out in Sect.~\ref{sec:spectroscopy}. 

\section{Discussion and conclusion}
\label{sec:discussion}

In this paper, we analyzed archival \sax\ and \rxte\ observations of \fu\ performed 
during the giant outburst of the source, which occurred in 1999. We investigated in particular 
the shape of the pulse profiles of the source at energies close to those of the cyclotron 
absorption features ($\sim12$, 24, 36, and 48\,keV).  At these energies, we measured a
peculiar phase shift of the pulse profiles for the first time by using a cross-correlation 
method. 

To interpret these findings, we created a geometrical model for the emission from an accretion 
column of a neutron star that considers the most relevant relativistic effects. 
The intrinsic beam pattern was calculated by assuming a parameterization for the emission 
with both a downward and an upward 
beam, which mimics the presence of the radiation advected in the column by  
the in-falling accreting material and the effect of resonant scattering. 
This model permits the reconstruction of 
the asymptotic beam patterns from \fu\ as perceived by a distant observer and thus the calculation of 
the simulated pulse profiles from the source by adopting the geometrical parameters determined by \citet{sasaki2011}.

Our method reproduced the general shape of the observed pulse profiles 
from \fu\ relatively well and the properties of their energy-dependent phase shifts described 
in Sect.~\ref{sec:phaselag}. 
According to our interpretation, these phase shifts can stem from an energy-dependent
beaming of the radiation, which we modeled as a variable 
contribution of the upward directed beam, whose origin can be ascribed to the resonant 
nature of the scattering cross-section in a strong magnetic field.
The modeled pulse profiles get distorted at the energies corresponding to the CRSFs, creating a negative 
shift in phase that is in fairly good agreement with what is inferred from the observations. 
With the adopted system geometry, we were also able to naturally explain the  
phase-dependent measured properties of the absorption lines.

Although we adopted a phenomenological description of the radiation emerging from 
the accretion column  of an HMXB, our calculations revealed the key role of an energy-dependent beaming
for the pulse profile formation mechanism. Based on this idea, we were able to point out a
plausible explanation for the energy-dependent phase shifts that we discovered. 
A more complex approach, which will solve the radiation 
transport problem using the actual cross sections and a realistic column velocity profile, will
provide a self consistent model of the spectral and timing characteristics of the XRB X-ray emission.

Significant changes of the pulse profiles with the energy, especially close to  
the resonances of the cyclotron scattering cross-section, have been reported so far 
in a number of different XRBPs \citep[e.g., V\,0332$+$53][]{tsygankov2006}. 
Applying the analysis method developed here to these other sources 
is ongoing and will be published elsewhere.

\section*{Acknowledgements}
This research made use of data obtained through the High Energy Astrophysics Science Archive 
Research Center Online Service, provided by the NASA/Goddard Space Flight Center and the archive 
of \sax data available at the ASDC of ASI.  C.~F. thanks
the research groups of IAA-T\"ubingen and Dr. Remeis-Sternwarte in Bamberg for developing and 
making available useful scripts to analyze \rxte\ data and A. Segreto for providing the software
to perform phase resolved analysis of the \sax data. He also acknowledges illuminating 
discussions with J. Wilms and S. Paltani on the timing techniques, and the help of R. Doroshenko 
for the reduction of some \sax data. We thank ISSI for hosting and financing International
group meetings on modeling and observations of cyclotron lines in High Mass X-ray binaries.

\bibliographystyle{aa}
\bibliography{lags,notpublished}

\begin{thebibliography}{61}
\expandafter\ifx\csname natexlab\endcsname\relax\def\natexlab#1{#1}\fi

\bibitem[{Araya-G{\'o}chez \& Harding(2000)}]{araya2000}
Araya-G{\'o}chez, R.~A. \& Harding, A.~K. 2000, \apj, 544, 1067

\bibitem[{Basko \& Sunyaev(1976)}]{basko1976}
Basko, M.~M. \& Sunyaev, R.~A. 1976, Royal Astronomical Society, 175, 395

\bibitem[{Becker(1998)}]{becker1998}
Becker, P.~A. 1998, Astrophysical Journal v.498, 498, 790

\bibitem[{Becker \& Wolff(2007)}]{becker2007}
Becker, P.~A. \& Wolff, M.~T. 2007, The Astrophysical Journal, 654, 435

\bibitem[{Beloborodov(2002)}]{beloborodov2002}
Beloborodov, A.~M. 2002, The Astrophysical Journal, 566, L85

\bibitem[{Blum \& Kraus(2000)}]{blum2000}
Blum, S. \& Kraus, U. 2000, The Astrophysical Journal, 529, 968

\bibitem[{Boella {et~al.}(1997{\natexlab{a}})Boella, Butler, Perola, Piro,
  Scarsi, \& Bleeker}]{sax}
Boella, G., Butler, R.~C., Perola, G.~C., {et~al.} 1997{\natexlab{a}}, \aaps,
  122, 299

\bibitem[{Boella {et~al.}(1997{\natexlab{b}})Boella, Chiappetti, Conti,
  Cusumano, del Sordo, Rosa, Maccarone, Mineo, Molendi, Re, Sacco, \&
  Tripiciano}]{mecs}
Boella, G., Chiappetti, L., Conti, G., {et~al.} 1997{\natexlab{b}}, \aaps, 122,
  327

\bibitem[{Bradt {et~al.}(1993)Bradt, Rothschild, \& Swank}]{rxte}
Bradt, H.~V., Rothschild, R.~E., \& Swank, J.~H. 1993, Astronomy and
  Astrophysics Supplement Series (ISSN 0365-0138), 97, 355

\bibitem[{Caballero {et~al.}(2010)Caballero, Kraus, Santangelo, Sasaki, \&
  Kretschmar}]{caballero2010}
Caballero, I., Kraus, U., Santangelo, A., Sasaki, M., \& Kretschmar, P. 2010,
  arXiv, astro-ph/1012.3077v1

\bibitem[{Cominsky {et~al.}(1978)Cominsky, Clark, Li, Mayer, \&
  Rappaport}]{cominski1978}
Cominsky, L., Clark, G.~W., Li, F., Mayer, W., \& Rappaport, S. 1978, \nat,
  273, 367

\bibitem[{Davidson \& Ostriker(1973)}]{Davids73}
Davidson, K. \& Ostriker, J.~P. 1973, \apj, 179, 585

\bibitem[{Ferrigno {et~al.}(2009)Ferrigno, Becker, Segreto, Mineo, \&
  Santangelo}]{ferrigno2009}
Ferrigno, C., Becker, P.~A., Segreto, A., Mineo, T., \& Santangelo, A. 2009,
  Astronomy and Astrophysics, 498, 825

\bibitem[{Ferrigno {et~al.}(2007)Ferrigno, Segreto, Santangelo, Wilms,
  Kreykenbohm, Denis, \& Staubert}]{ferrigno2007}
Ferrigno, C., Segreto, A., Santangelo, A., {et~al.} 2007, A{\&}A, 462, 995

\bibitem[{Frank {et~al.}(2002)Frank, King, \& Raine}]{frank2002}
Frank, J., King, A., \& Raine, D.~J. 2002, Accretion Power in Astrophysics,
  iSBN: 0521620538

\bibitem[{Frontera {et~al.}(1997)Frontera, Costa, dal Fiume, Feroci, Nicastro,
  Orlandini, Palazzi, \& Zavattini}]{pds}
Frontera, F., Costa, E., dal Fiume, D., {et~al.} 1997, \aaps, 122, 357

\bibitem[{Giacconi {et~al.}(1971)Giacconi, Gursky, Kellogg, Schreier, \&
  Tananbaum}]{giac71}
Giacconi, R., Gursky, H., Kellogg, E., Schreier, E., \& Tananbaum, H. 1971,
  \apjl, 167, L67

\bibitem[{Heindl {et~al.}(1999)Heindl, Coburn, Gruber, Pelling, Rothschild,
  Wilms, Pottschmidt, \& Staubert}]{heindl1999}
Heindl, W.~A., Coburn, W., Gruber, D.~E., {et~al.} 1999, The Astrophysical
  Journal, 521, L49

\bibitem[{Heindl {et~al.}(2004)Heindl, Rothschild, Coburn, Staubert, Wilms,
  Kreykenbohm, \& Kretschmar}]{heindl2004}
Heindl, W.~A., Rothschild, R.~E., Coburn, W., {et~al.} 2004, X-RAY TIMING 2003:
  Rossie and Beyond. AIP Conference Proceedings, 714, 323

\bibitem[{Isenberg {et~al.}(1998)Isenberg, Lamb, \& Wang}]{isenberg1998}
Isenberg, M., Lamb, D.~Q., \& Wang, J. C.~L. 1998, The Astrophysical Journal,
  505, 688

\bibitem[{Jahoda {et~al.}(1996)Jahoda, Swank, Giles, Stark, Strohmayer, Zhang,
  \& Morgan}]{pca}
Jahoda, K., Swank, J.~H., Giles, A.~B., {et~al.} 1996, Proc. SPIE Vol. 2808,
  2808, 59

\bibitem[{Johns {et~al.}(1978)Johns, Koski, Canizares, McClintock, Rappaport,
  Clark, Cominsky, \& Li}]{johns1978}
Johns, M., Koski, A., Canizares, C., {et~al.} 1978, \iaucirc, 3171, 1

\bibitem[{Klochkov {et~al.}(2008)Klochkov, Santangelo, Staubert, \&
  Ferrigno}]{dima2008}
Klochkov, D., Santangelo, A., Staubert, R., \& Ferrigno, C. 2008, A{\&}A, 491,
  833

\bibitem[{Kraus(2001)}]{kraus2001}
Kraus, U. 2001, The Astrophysical Journal, 563, 289

\bibitem[{Kraus {et~al.}(1996)Kraus, Blum, Schulte, Ruder, \&
  Meszaros}]{kraus1996}
Kraus, U., Blum, S., Schulte, J., Ruder, H., \& Meszaros, P. 1996,
  Astrophysical Journal v.467, 467, 794

\bibitem[{Kraus {et~al.}(1989)Kraus, Herold, Maile, Nollert, \&
  Rebetzky}]{kraus1989}
Kraus, U., Herold, H., Maile, T., Nollert, H.-P., \& Rebetzky, A. 1989,
  Astronomy and Astrophysics (ISSN 0004-6361), 223, 246

\bibitem[{Kraus {et~al.}(1995)Kraus, Nollert, Ruder, \& Riffert}]{kraus1995}
Kraus, U., Nollert, H.-P., Ruder, H., \& Riffert, H. 1995, Astrophysical
  Journal v.450, 450, 763

\bibitem[{Kraus {et~al.}(2003)Kraus, Zahn, Weth, \& Ruder}]{kraus2003}
Kraus, U., Zahn, C., Weth, C., \& Ruder, H. 2003, The Astrophysical Journal

\bibitem[{Leahy(1990)}]{leahy1990}
Leahy, D.~A. 1990, Royal Astronomical Society, 242, 188

\bibitem[{Leahy(1991)}]{leahy1991}
Leahy, D.~A. 1991, Royal Astronomical Society, 251, 203

\bibitem[{Leahy(2003)}]{leahy2003}
Leahy, D.~A. 2003, The Astrophysical Journal, 596, 1131

\bibitem[{Leahy(2004{\natexlab{a}})}]{leahy2004a}
Leahy, D.~A. 2004{\natexlab{a}}, Monthly Notices RAS, 348, 932

\bibitem[{Leahy(2004{\natexlab{b}})}]{leahy2004b}
Leahy, D.~A. 2004{\natexlab{b}}, The Astrophysical Journal, 613, 517

\bibitem[{Leahy \& Li(1995)}]{leahy1995}
Leahy, D.~A. \& Li, L. 1995, Monthly Notices RAS, 277, 1177, (c) 1995 The Royal
  Astronomical Society

\bibitem[{Manzo {et~al.}(1997)Manzo, Giarrusso, Santangelo, Ciralli, Fazio,
  Piraino, \& Segreto}]{hp}
Manzo, G., Giarrusso, S., Santangelo, A., {et~al.} 1997, \aaps, 122, 341

\bibitem[{Meszaros \& Nagel(1985{\natexlab{a}})}]{meszaros1985a}
Meszaros, P. \& Nagel, W. 1985{\natexlab{a}}, Astrophysical Journal, 298, 147

\bibitem[{Meszaros \& Nagel(1985{\natexlab{b}})}]{meszaros1985b}
Meszaros, P. \& Nagel, W. 1985{\natexlab{b}}, Astrophysical Journal, 299, 138

\bibitem[{Meszaros \& Riffert(1988)}]{riffert1988b}
Meszaros, P. \& Riffert, H. 1988, Astrophysical Journal, 327, 712

\bibitem[{Mihara {et~al.}(2004)Mihara, Makishima, \& Nagase}]{mihara2004}
Mihara, T., Makishima, K., \& Nagase, F. 2004, \apj, 610, 390

\bibitem[{Misner {et~al.}(1973)Misner, Thorne, \& Wheeler}]{misner1973}
Misner, C.~W., Thorne, K.~S., \& Wheeler, J.~A. 1973, San Francisco: W.H.
  Freeman and Co.

\bibitem[{Nakajima {et~al.}(2006)Nakajima, Mihara, Makishima, \&
  Niko}]{nakajima2006}
Nakajima, M., Mihara, T., Makishima, K., \& Niko, H. 2006, The Astrophysical
  Journal, 646, 1125

\bibitem[{Negueruela \& Okazaki(2001)}]{negueruela2001a}
Negueruela, I. \& Okazaki, A.~T. 2001, A{\&}A, 369, 108

\bibitem[{Negueruela {et~al.}(2001)Negueruela, Okazaki, Fabregat, Coe, Munari,
  \& Tomov}]{negueruela2001b}
Negueruela, I., Okazaki, A.~T., Fabregat, J., {et~al.} 2001, A{\&}A, 369, 117

\bibitem[{Pringle \& Rees(1972)}]{Pringle72}
Pringle, J.~E. \& Rees, M.~J. 1972, \aap, 21, 1

\bibitem[{Rappaport {et~al.}(1978)Rappaport, Clark, Cominsky, Li, \&
  Joss}]{rappaport1978}
Rappaport, S., Clark, G.~W., Cominsky, L., Li, F., \& Joss, P.~C. 1978, \apjl,
  224, L1

\bibitem[{Riffert \& Meszaros(1988)}]{riffert1988a}
Riffert, H. \& Meszaros, P. 1988, Astrophysical Journal, 325, 207

\bibitem[{Riffert {et~al.}(1993)Riffert, Nollert, Kraus, \&
  Ruder}]{riffert1993}
Riffert, H., Nollert, H.-P., Kraus, U., \& Ruder, H. 1993, Astrophysical
  Journal, 406, 185

\bibitem[{Rothschild {et~al.}(1998)Rothschild, Blanco, Gruber, Heindl,
  MacDonald, Marsden, Pelling, Wayne, \& Hink}]{hexte}
Rothschild, R.~E., Blanco, P.~R., Gruber, D.~E., {et~al.} 1998, Astrophysical
  Journal v.496, 496, 538

\bibitem[{Santangelo {et~al.}(1999)Santangelo, Segreto, Giarrusso, dal Fiume,
  Orlandini, Parmar, Oosterbroek, Bulik, Mihara, Campana, Israel, \&
  Stella}]{santangelo1999}
Santangelo, A., Segreto, A., Giarrusso, S., {et~al.} 1999, \apjl, 523, L85

\bibitem[{Sasaki {et~al.}(2010)Sasaki, Klochkov, Kraus, Caballero, \&
  Santangelo}]{sasaki2010}
Sasaki, M., Klochkov, D., Kraus, U., Caballero, I., \& Santangelo, A. 2010,
  Astronomy and Astrophysics, 517, 8

\bibitem[{Sasaki {et~al.}(2011)Sasaki, M\"uller, Kraus, Ferrigno, \&
  Santangelo}]{sasaki2011}
Sasaki, M., M\"uller, D., Kraus, U., Ferrigno, C., \& Santangelo, A. 2011,
  Astronomy and Astrophysics, submitted

\bibitem[{Sch{\"o}nherr {et~al.}(2007)Sch{\"o}nherr, Wilms, Kretschmar,
  Kreykenbohm, Santangelo, Rothschild, Coburn, \& Staubert}]{gabi2007}
Sch{\"o}nherr, G., Wilms, J., Kretschmar, P., {et~al.} 2007, A{\&}A, 472, 353

\bibitem[{Tamura {et~al.}(1992)Tamura, Tsunemi, Kitamoto, Hayashida, \&
  Nagase}]{tamura1992}
Tamura, K., Tsunemi, H., Kitamoto, S., Hayashida, K., \& Nagase, F. 1992,
  Astrophysical Journal, 389, 676

\bibitem[{Tananbaum {et~al.}(1972)Tananbaum, Gursky, Kellogg, Levinson,
  Schreier, \& Giacconi}]{tananbaum1972}
Tananbaum, H., Gursky, H., Kellogg, E.~M., {et~al.} 1972, \apjl, 174, L143

\bibitem[{Tsygankov {et~al.}(2006)Tsygankov, Lutovinov, Churazov, \&
  Sunyaev}]{tsygankov2006}
Tsygankov, S.~S., Lutovinov, A.~A., Churazov, E.~M., \& Sunyaev, R.~A. 2006,
  Monthly Notices RAS, 371, 19

\bibitem[{Tsygankov {et~al.}(2007)Tsygankov, Lutovinov, Churazov, \&
  Sunyaev}]{russi2007}
Tsygankov, S.~S., Lutovinov, A.~A., Churazov, E.~M., \& Sunyaev, R.~A. 2007,
  Astronomy Letters, 33, 368, (c) 2007: Pleiades Publishing. Inc

\bibitem[{Tsygankov {et~al.}(2010)Tsygankov, Lutovinov, \&
  Serber}]{tsygankov2010}
Tsygankov, S.~S., Lutovinov, A.~A., \& Serber, A.~V. 2010, Monthly Notices RAS,
  401, 1628

\bibitem[{Wasserman \& Shapiro(1983)}]{wasserman1983}
Wasserman, I. \& Shapiro, S.~L. 1983, Astrophysical Journal, 265, 1036

\bibitem[{Wheaton {et~al.}(1979)Wheaton, Doty, Primini, Cooke, Dobson, Goldman,
  Hecht, Howe, Hoffman, \& Scheepmaker}]{wheaton1979}
Wheaton, W.~A., Doty, J.~P., Primini, F.~A., {et~al.} 1979, \nat, 282, 240

\bibitem[{White {et~al.}(1983)White, Swank, \& Holt}]{white1983}
White, N.~E., Swank, J.~H., \& Holt, S.~S. 1983, \apj, 270, 711

\bibitem[{Whitlock {et~al.}(1989)Whitlock, Roussel-Dupre, \&
  Priedhorsky}]{whitlock1989}
Whitlock, L., Roussel-Dupre, D., \& Priedhorsky, W. 1989, \apj, 338, 381

\end{thebibliography}

\appendix

\section{Calculation of the observed flux from an accretion column}
\label{sec:appendix}

\begin{figure}
  \begin{center}
\resizebox{\hsize}{!}{\includegraphics[angle=0]{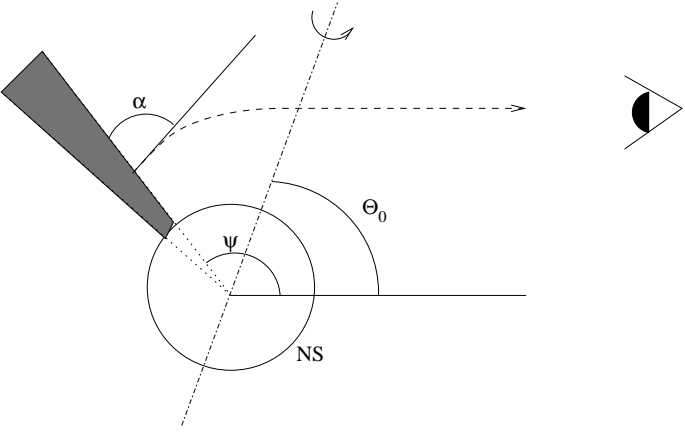}}
         \caption{The geometry of the systems (not in scale). The dashed line shows a photon trajectory from 
         the accretion column surface to the observer located at infinity, the dot-dashed line is the rotational axis of the NS.
         }
\label{fig:psi_alpha}
\end{center}
\end{figure}

We present a relativistic ray-tracing computations of the beam pattern emitted from the accretion column surface of a slowly rotating neutron star to model the phase lags observed in the X-ray pulse profiles of \fu.
We consider the geometrical properties of the column and the relativistic effects, i.e., light bending and the lensing effect in a Schwarzschild metric. Since the NS is slowly rotating, the relativistic travel time delay and the Doppler boosting have not been taken into account. The observer is located at positive infinity on the $z$-axis in the non-rotating observer frame with polar coordinates ($r, \theta, \phi$). The accretion column reference frame has the polar 
coordinates ($r^\prime, \theta^\prime, \phi^\prime$) with the $z^\prime$-axis coincident with the accretion column axis. 
See Fig. \ref{fig:psi_alpha} for the geometry of the systems. The deflection angle of a photon emitted from the accretion column surface is $\psi$, while $\alpha$ indicates the angle between the local radial direction and the photon direction at the emission point. 

\subsection{Light bending}

The geodetic equation for an emitting position of a photon depends on the impact parameter, $b$ \citep{misner1973}, 
which is defined as
\begin{equation}
b = \frac{r \sin \alpha}{\sqrt{1-r_\mathrm{s}/r}} ,
\label{eq:b}
\end{equation}
where $r$ is the radial coordinate of the emitting surface element,
and $r_\mathrm{s} = 2GM/c^2$ is defined as the Schwarzschild radius.

The relation between $\alpha$ and the photon deflection angle, $\psi$, i.e., the light bending, can be directly obtained using the approximated null geodetic equation, strictly valid for $\alpha<\pi/2$ \citep[see][for more details]{beloborodov2002}: 
\begin{equation}
1-\cos\alpha = (1-\cos\psi)\left(1-\frac{r_\mathrm{s}}{r}\right) .
\label{eq:belo}
\end{equation}
 For each column surface element at a given column orientation, $\psi_\mathrm{obs}$,  
(see Eq.~\ref{eq:psi_obs}), we determine the angle, $\alpha$, and ,therefore, $\psi$, at which the emitted radiation  
reaches the observer (the only trajectories considered for the 
pulse profile modeling\footnote{We neglect values of the angle $\psi$
such that  $\psi > \pi$. For these values multiple images of 
an emitting point could be formed. This is marginally relevant only for an 
accretion column located behind the NS, and thus not for \fu\ 
(see Sect.~\ref{sec:model}).}).  

\begin{figure}
   \begin{center}
	\resizebox{\hsize}{!}{\includegraphics[angle=0]{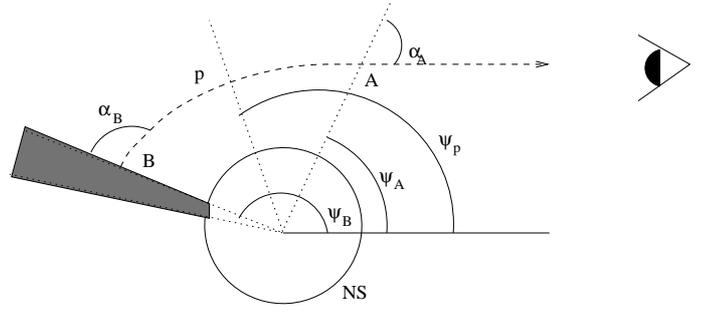}}
          \caption{Schematic representation of a photon trajectory 
           (dashed line) from the accretion column surface to the distant 
           observer for trajectories with a
          turning point (not in scale). The emission point is $B$ and the angle to be computed is  
          $\psi_B$. $A$ is the point at which the trajectory reaches for the 
          second time the radial coordinate at which it originated. $p$ indicates 
          the location of the trajectory's periastron.
         }
\label{fig:nuova}
\end{center}
\end{figure}

For photons emitted towards the NS surface ($\alpha > \pi/2$), 
the allowed $\alpha$ values can be estimated by imposing that the trajectory  
not be swallowed by the relativistic photosphere of the compact object. This translates into the relation
$\pi/2<\alpha<\alpha_\mathrm{max}$, where 
\begin{equation}
\alpha_\mathrm{max}=\pi - \sin^{-1} \left( \frac{3}{2} \sqrt{3 \left[1- 
\frac{r_\mathrm{s}}{r} \right]} \frac{r_\mathrm{s}}{r}  \right),
\end{equation}
and $r$ is the radial coordinate at which the 
photon is emitted from the column (e.g., point $B$ in Fig.~\ref{fig:nuova}). These trajectories have a ``turning point'' and thus
possess a periastron; it is then convenient to express all relevant quantities in terms of the periastron distance, $p$. 
For $\alpha_\mathrm{max}>\alpha_B>\pi/2$, we first compute the impact  
parameter using Eq.~(\ref{eq:b}), and then estimate the value $p$ of the periastron by 
taking the largest real solution of the equation $ p^3-b^2p+b^2r_\mathrm{s}=0$. 
This equation is obtained by setting $\alpha=\pi/2$ and $r=p$ in Eq.~(\ref{eq:b}),  
as $\alpha_p=\pi/2$ at the periastron. Using Eq.~(\ref{eq:belo}), we find  
$\psi_p=\cos^{-1} (1 - 1/\left[1 - r_\mathrm{s}/ p  \right])$ . The  
photon trajectory again reaches the radial coordinate $r$ at point  
$A$ (see Fig.~\ref{fig:nuova}), where the angle formed by the photon  
with the local radial direction is $\alpha_A=\pi - \alpha_B$.   
The angle $\psi_A$ can now be computed from Eq.~(\ref{eq:belo}). Finally, from  
simple symmetry considerations,
\begin{equation}
\psi_B = 2 \psi_p - \psi_A\,. 
\end{equation}
This characterizes the considered trajectory fully, since the impact parameter for $\psi_A$  has the same value as for the deflection angle at the emission point, $\psi_B$. 

For each photon-emitting point we can now follow the photon 
trajectory to verify if it is absorbed by the NS or accretion column surface.  
For $\alpha<\pi/2$, the trajectory can be directly computed using the following equation \citep{beloborodov2002}: 
\begin{equation}
\widehat{r}(\widehat{\psi}) = \left[ \frac { r_\mathrm{s}^2 ( 1- \cos \widehat{\psi})^2}{4 (1+\cos \widehat{\psi} )^2} 
+ \frac{b^2}{\sin^2\widehat{\psi}}\right]^2 -
\frac{r_\mathrm{s}^2 (1-\cos\widehat{\psi})}{2(1-\cos\widehat{\psi})}\,, 
\end{equation}
where $\widehat{\psi}$ values are in the range (0,$\psi$). 

For $\alpha>\pi/2$, we use Eq.~(\ref{eq:b})
to link $\widehat{\alpha}$ to $\widehat{r}$ along the trajectory (the impact parameter $b$ is constant in each geodesic). 
The corresponding value of $\widehat{\psi}$ is found with the 
method outlined above.
The trajectories that hit the NS or intersect one of the optically thick accretion columns are 
excluded from the computation of the source flux measured 
at the observer's location. 

\subsection{Lensing and red shift}

We now compute the flux emitted by the column and measured by a distant observer. 
The flux emitted by a surface element from the accretion column, when it reaches the observed plane normal to the direction of the line of sight, is given by $\mathrm{d} F_\nu = I_\nu \mathrm{d}\Omega$, where the solid angle can be expressed in terms of the impact parameter \citep{misner1973}:
 \begin{equation}
 \mathrm{d} F_\nu = \frac{I_\nu (b,\phi) b \mathrm{d} b \mathrm{d} \phi }{D^2} . 
 \label{eq:flux}
 \end{equation}
Here, $\phi$ is the azimuthal angle around the line of sight of the plane containing the trajectory of the photon, and $I_\nu(b,\phi)$ is the differential intensity of the radiation emitted by the surface element when it reaches  
the observed plane at a distance $D$. 

The impact factor $b$
depends on $r$ and $\alpha$, but not on $\phi$. The angle $\alpha$ 
can be expressed as a function of  $r$ and $\psi$, as shown in the previous section. 
We further note that the differential in Eq.~(\ref{eq:flux}) can be expressed as
$\mathrm{d}b=(\mathrm{d}b/\mathrm{d}r)_\psi \mathrm{d}r + (\mathrm{d}b/\mathrm{d}\psi)_r \mathrm{d}\psi$, and
is computed along a line on a geometrical cone at constant $\phi$, which is by definition a radial line. 
Therefore we can simplify the differential and write $\mathrm{d}b=(\mathrm{d}b/\mathrm{d}r)_\psi \mathrm{d}r $, since $\psi$ is constant along a radial line.

Up to this point we expressed the equations in the observer's reference frame defined above. 
To integrate the emission on the column surface, we transform the integration variables to the column's frame
to use quantities directly related to the accretion 
column. The Jacobian of 
the coordinate transformation involves only the variables
$r$ (constant) and $\phi$, and is:
$\left|\mathrm{d}\phi/\mathrm{d}\phi^\prime\right|$;   
therefore, Eq.~(\ref{eq:flux}) can be written as
 \begin{equation}
 \mathrm{d} F_\nu = \frac{I_\nu (\alpha,\phi^\prime,r) b(\alpha,r) \frac{\mathrm{d} b}
 {\mathrm{d} r} \left|  \frac{\mathrm{d}\phi}{\mathrm{d}\phi^\prime} \right|  \mathrm{d}
 \phi^\prime  \mathrm{d}r}{D^2}. 
 \label{eq:flux_col}
 \end{equation}
We still need to express $I_\nu$ in the system of reference of the 
column to account for the gravitational red-shift. Since $I/\nu^3$ is a Lorentz invariant, we can write
 \begin{equation}
I_\nu = \left( \frac {\nu_0 }{ \nu} \right)^3 I_{0,\nu_0} = \left( 1- 
\frac{r_\mathrm{s}}{r}\right)^{3/2} I_{0,\nu_0}. 
\label{eq:int}
\end{equation}

As already described in Sect.~\ref{sec:model}, knowing the exact functional shape of $I_{0,\nu_0}(r,\alpha)$ would require 
a detailed treatment of the cyclotron scattering process in a strong magnetic field 
($B\sim10^{12}$\,G) and of the relativistic beaming caused by the bulk motion of the 
plasma in the accretion column above the surface of the NS. 
This is outside the scope of the present paper. 
Instead, we aim at reproducing the beam pattern in the energy range of interest ($>10$\,keV)  found 
by using the pulse decomposition method described in \citet{sasaki2011}. For this purpose, we use a 
combination of two Gaussians, one pointed downwards and one upwards characterized 
by different amplitudes: 
\begin{equation}
I_0(\alpha,r) = \sum_{i=1}^{2} N_i \exp \left( -\frac{1}{2} \left( \frac{\alpha - 
\alpha_i}{\sigma_i}\right)^2\right). 
\label{eq:ialpha}
\end{equation}
In the equation above, $\alpha_1=150\gra$, $\alpha_2=0\gra$, $\sigma_1=45\gra$, 
$\sigma_2=45\gra$, $N_1=100$, and $N_2$ assumed values from 0 to 100 to mimic 
a variable contribution from the cyclotron scattered radiation (See also 
Sect.~\ref{sec:model}). 

The total flux emitted by the column is finally estimated from  
Eqs.~(\ref{eq:flux_col}), (\ref{eq:int}), and (\ref{eq:ialpha}) 
performing a Monte-Carlo integration with $\phi^\prime \in (0,2\pi)$, and 
$r\in(r_\mathrm{NS},r_\mathrm{max})$. The maximum height for 
the accretion column is equal to $r_\mathrm{max}-r_\mathrm{NS}$. 
We did not include in the total flux the contribution of the photons that hit the NS surface and/or 
are absorbed by the column. 

\begin{figure}
  \begin{center}
\resizebox{\hsize}{!}{\includegraphics[angle=0]{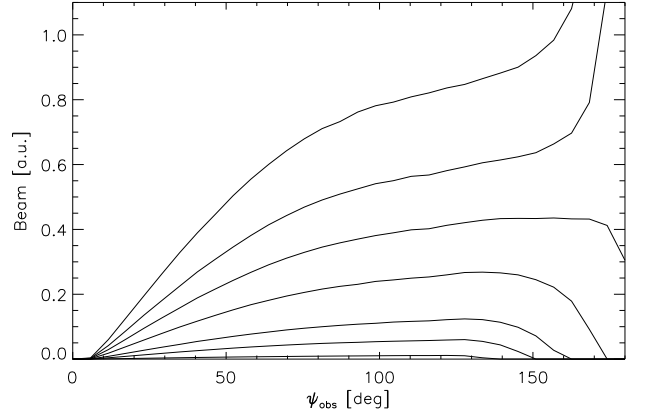}}
         \caption{Column beam pattern seen by a distant observer and calculated 
         for an isotropic emission in the case of a NS with a radius $r=3r_\mathrm{S}$. The accretion column 
         is assumed to have an half-aperture of 5 degrees, and an heights of (from bottom to top) 3.03, 3.15, 
         3.3, 3.6, 3.9, 4.2, and 4.6 $r_\mathrm{S}$. The angle $\psi$ is measured from the line of 
         sight to the axis of the column.}
\label{fig:beams_const}
\end{center}
\end{figure}

To verify the consistency of our results with similar calculations published before, 
we show in Fig.~\ref{fig:beams_const} the beam patterns obtained from 
an isotropic emission on the lateral surface of an accretion column 
with the same configuration adopted in Fig.~3 of \citet{riffert1988a}. 
We found fairly good agreement between our results and those published by these authors. 
There is only a negligible discrepancy in the estimated fluxes 
of an accretion column extending up to 3.9 gravitational radii above the NS surface. Exact 
calculations predict that the critical height of a column for its total obscuration in the antipodal 
position should be 3.94$r_\mathrm{s}$. With 
the approximation used in our calculations, we obtained a value of 3.85$r_\mathrm{s}$. 
This discrepancy can only be
relevant for a height of the column that is close to this critical value 
and for angles close to the antipodal position.  Our approximation thus gives 
a satisfiable result for the accuracy required in the present work.  

\end{document}